\documentclass{article}
\usepackage[utf8]{inputenc}
\usepackage{amssymb, amsmath, amsthm, amsfonts}
\usepackage[onehalfspacing]{setspace}
\usepackage{graphicx}
\usepackage[english]{babel}
\usepackage{tikz}
\usepackage{caption}
\usepackage{subcaption}
\usepackage{float}
\usepackage{natbib}

\title{A Trimming Estimator for the Latent-Diffusion-Observed-Adoption Model}
\author{L.S. Sanna Stephan }
\date{September 2023}

\begin{document}

\maketitle

\section{Introduction}

Diffusion in networks has been vastly studied in biology and Computer Science. Yet recently it has increasingly been acknowledged that also economic agents exchange goods, services or assets in social networks and that it is highly important and interesting to estimate at which rate they do so. 
A key challenge here is that economic interaction in social networks is oftentimes unobserved. 
\\
\noindent We can tackle such a situation assuming that economic events which we spot are but a signal of the process that jointly determines an observed outcome and an unobserved state variable. Yet, fitting such a model to the data is challenging  due to the high dimensionality: at each point in time, there exists a system of equations relating all the unobserved individual state variables to one another, and the dynamics of this system are governed by the network. With social networks typically being dense and large, we thus have a gigantic number of scenarios that could have given rise to the data at hand. This is problematic in particular if the time horizon we wish to consider is large.
\\ 
\noindent Due to the nonlinear nature of network diffusion models, Maximum Likelihood (ML) estimation would be the preferred estimation method. However, setting up the log likelihood function is computationally infeasible for larger time horizons. 
The gigantesque number of possible realizations of the latent diffusion process that could have generated the observed adoption pattern makes it impossible to establish the probability to observe the given data by applying the law of total probability (i.e.\ by integrating out the latent process).
To ensure tractability one needs to either restrict the model time horizon or to use a dimension reduction strategy. In this paper, I conduct a Monte Carlo study in order to compare these two alternatives. This shows that the ``Trimming estimator", derived in the present chapter, outperforms the previously used estimator that restrict the number of rounds of network interaction that can take place in most cases (precisely, whenever the sample includes a sufficient number of villages for which trimming impacts no more than one-third of the individuals that are eligible to be trimmed). The ``Trimming estimator" is thereafter applied to four real villages from the \citet{BCDJ} study. 
\\ \noindent
I propose a simple trimming rule that imposes restrictions on the number of permissible results of the unobservable network interaction and compare it to the estimate that results from restricting the time horizon modeled. Fast and flexible, the algorithm can be used to estimate a class of network models that comprises an observed outcome and an unobserved state variable. The strategy is easily implemented and oftentimes succeeds in selecting the interaction outcomes with the largest probability mass. I also investigate how features of the network affect performance of the algorithm.\\
\noindent 
I demonstrate, that the MLE obtained from maximizing the approximate log likelihood function quickly converges to the exact MLE as the trimming value increases. 
\noindent 
I proceed by establishing the (approximate) log likelihood function and finding its maximum by means of grid search. This entails certain advantages. As a byproduct we obtain the log likelihood function which, plotted over the grid, can provide us with additional insights (e.g.\ I can detect local maxima, areas of weak identification or illustrate the covariance of the parameters). Confidence sets can be constructed straightforwardly using the Likelihood Ratio (LR) test and corner solutions can be dealt with. \\
\noindent The application considered stems from the seminal study \citet{BCDJ}, in which the authors investigate the diffusion of micro-finance in Indian village networks. Linked agents repeatedly exchange new information, the information status being unobserved. Upon being newly informed, they can make an observable borrowing choice. We use eleven of the village networks from this data set in our simulation study. Nonetheless, it is important to emphasize that the applicability of our algorithm goes well beyond this particular model. 
A model in which an observed outcome depends on an unobserved status lends itself to various applications in  economics. Instead of information, one  could model technology spill-over, where we observe the adoption of new technologies, but for those that abstain from it we do not know whether this is by deliberate choice or by lack of knowledge. Alternatively, one can estimate a model in which the probability of (observed) micro-business default depends on (unobserved) mutual lending among friends.
\\ \noindent 
In recent years, the amount of data on (action taking place through) social networks that is available has increased drastically, 
providing new exciting research possibilities, but also serious challenges for econometricians. What shies away many applied researchers from fully embracing this opportunity and using MLE in settings where it provides clear advantages is often the lack of tractable, easily implementable algorithms. While established software packages exist for common estimation procedures, these typically work very well for specific tasks, but lack the flexibility required by network models. They are not fully efficient in terms of speed and provide little opportunity to deal with large dimensionality of the data. Research into methodological approaches facilitating ML estimation in such models thus has the potential to be beneficial. In the current paper, I make a contribution to the this new strand of research.\\
\noindent The paper is organized as follows: 
Section 2 provides an overview of the model. In section 3 we highlight the challenges of ML estimation in this model and introduce the basic trimming procedure. Section 4 outlines the Monte Carlo study. Section 5 provides and analyses the Monte Carlo results. Section 6 contains the results from applying the trimming estimator to four real villages using actual adoption data. Section 7 concludes. The Appendices highlight how features of the network impact the algorithm's performance and the shape of the error curve. 

\section{The Model}

In this and the following sections, we use the terminology of the concrete application despite our algorithm being applicable to many research questions. \\ \noindent
We dispose over a sample of villages, indexed by $v=1,...,V$. 
\\ \noindent $N_v$ individuals in village $v$ are linked in a network, represented by the network adjacency matrix $G_v$. While villagers $i=1,...,N_v$ entertain relationships among themselves, they are supposed not to interfere with agents living in other villages. 
Consequently, to ease notation, in the following we drop the subscript $v$ and we outline the model with an exemplary village. \\ \noindent
The symmetric matrix $G$ comprises binary variables indicating the presence (absence) of a link. 
If the entry in row $i$, column $j$ takes the value of one (i.e. $g_{ij}=g_{ji}=1$), we denominate $i$ and $j$ as “neighbors" or “friends". An organization enters the community and starts providing a new technology to its members. 
In period $t=0$, they advertise their innovation  to a subset of them (referred to as the “information injection points" or “IPs") and thereafter rely on word-of-mouth marketing.   \\
\noindent
In each subsequent period ($t=1,...,4$), two processes take place: first, newly informed individuals face the choice of whether or not to adopt the technology, second, informed individuals can instruct their neighbors about the novel opportunity. \\ \noindent 
The process is modeled using the random matrices $Y$ and $S$, containing dummy variables for respectively the participation and information statuses of each of the villagers at each point in time. 
Using the first subscript for individuals and the second subscript for time periods, thus $Y_{it}=1$ ($S_{it}=1$) (i.e.\ the row $i$, column $t$ element of $Y$ ($S$) displays the value one) implies that $i$ participates (is informed) in period $t$. 
The distributions of the random matrices $Y$ and $S$ depend on the village network ($G$) and the information injection $s_{0}$).  
\\
\noindent
For each village, we observe the network ($G$), the information initialization $s_{0}$ and the individual participation decisions over time (i.e. one realization of $Y$). The information statuses of all inhabitants but the IPs are generally unobserved. In particular, there usually exist various realizations of $S$ that are in accordance with the data at hand. \\
\noindent The following assumptions are made: \\

\indent \textbf{Assumption 1:} Independence across villages \\
Villages can be treated as independent entities. In particular, there is no link between any two agents that reside in different communities. \\
\\
\indent \textbf{Assumption 2:} Exogenous network \\
The network is exogenous and observed. Measurement error is negligible. 
\\ \\
\indent \textbf{Assumption 3:} Timing \\
Each period consists of two processes: First, newly informed individuals decide upon participation, second, information is exchanged.
This implies that $Y$ is of dimension $N \times 4$ while $S$ is of dimension $N \times 3$: since information is exchanged after the participation decisions, modeling the fourth period's information exchange is redundant.
\\ \\
\indent \textbf{Assumption 4:} Information is a pre-condition for participation \\ \\
\indent \textbf{Assumption 5:} Participation is a one-time opportunity \\ \noindent
Each period only \underline{newly informed} individuals face the participation decision. Having opted in (out), the respective individual will thereafter stay in the set of participants (non-participants) forever. \\
\\
\indent \textbf{Assumption 6:} Distributional assumption \\
Conditional on being newly informed, the random variables $ Y_{it}
\hspace{0,15cm} \forall i=1,...,N; \hspace{0,15cm} t=1,...,4$ are i.i.d. and follow a Bernoulli distribution. The probability to participate conditional on being newly informed is $p$. \\ \\
\indent \textbf{Assumption 7:} Information is never forgotten \\
Individuals can only switch their information status once (from being uninformed ($S_{i(t-1)}=0$) to being informed ($S_{it}=1$)). Any informed individual remains potentially informing her neighbors every period.  
\\ \\
\indent \textbf{Assumption 8:} Information exchange \\
In each period, informed individuals may transmit the information to their neighbors. On average, they do so with probability $q$. Transmitting is independent across all links.  \\ \noindent
Let the random variable $I_{(ij)t}$ denote the indicator that shows that individual $i$ has sent the information to individual $j$ in period $t$. 
Conditional on $i$ being a sender, the variables $I_{(ij)t} \hspace{0,1cm} \forall i\neq j; t=1,...,3 $ are i.i.d. and follow a Bernoulli distribution with $E[I_{(ij)t}]=q$. 
\\ \\ \noindent
The aim is to estimate \\
\indent (i) the individual's probability to participate if informed ($p$) and \\
\indent (ii) individual's probability to share information with acquaintances ($q$)

\section{Establishing the (approximate) Log Likelihood function}

\noindent Using the notation introduced above, 
\begin{equation}
\label{like}
\ell(p,q) =
Pr\big(Y_1=y_1,...,Y_T=y_T|S_0=s_0,G,p,q\big)
\end{equation}
is the probability to observe the actual data at hand, given the (nonrandom) village network, the parameters and the information initiation and
\begin{equation}
\label{Like}
\mathcal{L}(p,q)=
\log \hspace{.1cm} \Big(
Pr\big(Y_1=y_1,...,Y_T=y_T|S_0=s_0,G,p,q\big)
\Big)
\end{equation}
is the village log likelihood function. 
\\ \noindent 
However, the probability to observe any particular outcome pattern depends on the (unobserved) information propagation.\footnote{Note that knowing the probability of (not) observing somebody as a participant depends on her chances to receive the information which in turn depends on which of her neighbors are informed} In the following, we use “information scenario" to denote a unique sequence $S_1=s_1,...,S_{(T-1)}=s_{(T-1)}$. The set $\mathbb{S}$ comprises all possible information scenarios. Then the log likelihood function is (by the law of total probability)
\begin{equation}
\label{Lmain3}
=
\log \Bigg(
\displaystyle \sum_{s \in  \mathbb{S}
}
Pr \Big( Y_{1:T}=y_{1:T}, S_{1:(T-1)} \in s| S_0=s_0, G, p,q \Big) \Bigg)
\end{equation}
\noindent 
The joint probability of the outcome matrix and the information status matrix can, by iterated conditioning, conveniently be factorised into four time period specific factors, which in turn are a product of $N$ individual specific factors. Denoting by $l_t$ the likelihood up to and including period $t$ and by $\Delta \ell_t$ the change in the likelihood function as we move from period $t-1$ to period $t$, we can rewrite \eqref{like} as

\begin{equation} 
\label{factors}
\ell
= \underbrace{
\underbrace{
\underbrace{
\displaystyle \sum_{\tau=1}^{\# {\mathbb{S}} }
\ell_1 \times 
\Delta 
\ell_{2(\tau)} }_{\ell_2}
 \Delta
\ell_{3( \tau)}}_{\ell_3}
 \Delta
\ell_{4( \tau)}}_{\ell_4}
\hspace{1cm}
\Delta \ell_{t (\tau)}= \prod_{i=1}^N 
\Delta \ell_{it(\tau)}
\end{equation} 

\noindent As we shall see, for any particular information scenario, the factors appearing in \eqref{factors} are straightforward to evaluate and thus the joint probability of the outcome and that particular scenario are easily computed. Applying the law of total probability then amounts to summing up these joint probabilities over all possible information scenarios. The challenge stems from the fact that the cardinality of $\mathbb{S}$ quickly becomes prohibitively large.  

\noindent
Individual $i$'s contribution to the village log likelihood function depends on any other individual only insofar as that $i$ may newly receive the information from her. Let
\begin{equation}
r_{it}=P(S_{it}=1|S_{i(t-1)}=0, S_{t-1}=s_{t-1},G,p,q)
\end{equation}
denote $i$'s probability to newly receive the information (from any of her neighbors) in period $t$. Using $r_{i(t-1)}$, we can categorize villagers at any point in time into four groups according to what they contribute to $\ell$ at this point in time (that is, according to $\Delta \ell_{it}$): 
\begin{itemize}
\item {\bf former participants:} $\Delta \ell_{it}=1$ \\
These individuals are certain to remain (informed) participants until the end of the modeling horizon.
\item {\bf new participants:} $\Delta \ell_{it}=r_{i(t-1)}p$
\\
These individuals are observed to experience a switch in their outcome, which (according to the modeling assumptions) must be preceded by a switch in their information status. 
\item {\bf non-participants out of reach}
$\Delta \ell_{it}=1$\\
These individuals are too far away from any IP to possibly have received the information and thus certain to be uninformed.
\item {\bf potentially informed individuals (PIIs) } \\
Sufficiently close to an IP to potentially receive the information, they yet do not indicate (by participating) that they have done so.
As a consequence, while the likelihood contribution for the three groups above is unequivocal, for the PIIs, three scenarios (each leading to a specific likelihood contribution) are
 indistinguishable in terms of the empirical evidence that they generate.
\begin{itemize}
\item {\bf scenario A: newly informed PIIs} 
$\Delta \ell_{it}= P(A_{it})=
r_{i(t-1)} (1-p)$
\\
This group has received the information in the last exchange and opts out this period.
\item {\bf scenario B: uninformed PIIs }
$\Delta \ell_{it}=P(B_{it})=
(1-r_{i(t-1)})$
\\
They entered last period's information exchange uninformed and also did not receive the information. 
\item {\bf scenario C: previously informed PIIs }
$\Delta \ell_{it}=P(C_{it})=
1$
\\
Since they obtained the information and opted out in a previous period they will thus abstain from participation forever. 
\end{itemize}
\end{itemize}
We can thus see that the PIIs are the origin of the dimensionality problem: without them, there would be only one information status matrix that could have generated the data. 
PIIs introduce a twofold challenge. First, their own information status cannot be derived from the data and hence their own contribution to the likelihood function is scenario specific. Second, whether or not they enter any particular period informed impacts all their neighbours' (and later on their neighbours' neighbours') information reception probabilities, implying that these individuals' likelihood contributions become scenario specific as well. As the cardinality of $\mathbb{S}$ increases exponentially in the number of PIIs, hence \eqref{Like} cannot be computed. \\
\noindent Due to the interdependence described above, likelihood contributions being scenario specific implies that $i's$ contribution depends on the previous information statuses of all her neighbours as well as on her own current status.  \\
\noindent
As claimed above, for any particular information scenario, the individual likelihood contributions can be calculated straightforwardly: first, the individual information reception probabilities $r_{it}$ are computed, then these are plugged into the formulas above, where for PIIs we choose formula A, B or C, depending on her own past and current status.  
Since $i$ is informed as long as {\it any} of her neighbors sends the information, we can use the counter-factual probability (nobody informs $i$) to calculate 
\begin{equation}
r_{it}=1-\prod_j \underbrace{ (1-g_{ij}qs_{j(t-1)})}_{ 
\mbox{ \footnotesize
\begin{tabular}{c}
if $j$ is an informed neighbor of $i$,\\
she does not send the information
\end{tabular}} 
}
\end{equation}   
\noindent
From period two onward, different information status vectors from the last period lead to different PIIs and different probabilities of any individual to receive the information in the current period, thus $r_{it}$ needs to be computed for each $S_{(t-1)} \in \mathbb{S}$.
\\ \noindent 
The origin of the dimensionality problem will also be the point of attack for our algorithm since a {\bf restriction of the number of PIIs} directly translates into a pronounced reduction of information scenarios. \\
\noindent For any PII who enters a time period uninformed, the scenarios A: “newly informed and opted out" and B: “uninformed" are observationally equivalent. The basic idea of our trimming procedure is to, for a certain number of PIIs, restrict the log likelihood function to include only one of these two cases.  
\\ \noindent We illustrate the dimension reduction strategy using two toy villages.
The first village consists of six individuals one of which is the IP. Even in this very small and rather sparse network, three rounds of information exchanges result in 92 information propagation possibilities.
\begin{figure}[ht!]
    \centering
    \caption{Village Graph 1}    
\begin{tikzpicture}
\path
(0,0) node (1) [shape=circle,draw, inner sep=2,outer sep=.5] {1 }
(0,-.65) node (ip) [shape=circle,inner sep=.5,outer sep=0.5] {IP }
(2,.5) node (2) [shape=circle,draw, inner sep=2,outer sep=.5] {2 }
(2,-.5) node (3) [shape=circle,draw, inner sep=2,outer sep=.5] {3 }
(5,.5) node (4) [shape=circle,draw, inner sep=2,outer sep=.5] {4 }
(5,-.5) node (5) [shape=circle,draw, inner sep=2,outer sep=.5] {5 }
(7,0) node (6) [shape=circle,draw, inner sep=2,outer sep=.5] {6 }
;
\draw [dashed] (1)--(2);
\draw [dashed] (1)--(3);
\draw [dashed] (2)--(4);
\draw [dashed] (3)--(5);
\draw [dashed] (4)--(6);
\draw [dashed] (6)--(5);
\draw [dashed] (3)--(4);
\end{tikzpicture}
\end{figure}
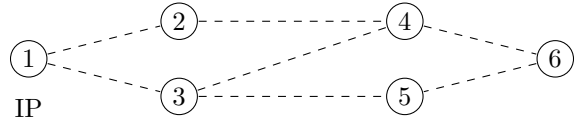

\begin{figure}[ht!] \centering 
\caption{Scenario Equivalence Thresholds for PIIs With 1 to 4 Informed Neighbours}
\label{thresh}
\begin{tikzpicture}[scale=5] 
\path
(1.075,0.9) node (1) [shape=circle, draw, inner sep=2,outer sep=.5, red] { 1 } 
(1.5,0.9) node (1) [shape=circle, inner sep=2,outer sep=.5] { $
p=
2- \frac{1}{q}
$ } 
(1.075,0.7) node (1) [shape=circle, draw, inner sep=2,outer sep=.5, orange] { 2 }
(1.55,0.7) node (1) [shape=circle, inner sep=2,outer sep=.5] { $
p= 
2- \frac{1}{(2q-q^2)}
$ }
(1.075,0.5) node (1) [shape=circle, draw, inner sep=2,outer sep=.5, cyan] { 3 }
(1.55,0.5) node (1) [shape=circle, inner sep=2,outer sep=.5] { $
p=
2- \frac{1}{(3q-3q^2+q^3)}
$ }
(1.075,0.3) node (1) [shape=circle, draw, inner sep=2,outer sep=.5, green] { 4 }
(1.55,0.3) node (1) [shape=circle, inner sep=2,outer sep=.5] { $
p= 
2- \frac{1}{(4q-6q^2+4q^3-q^4)}
$ }
;
\draw[dashed,lightgray] (0,0.1) -- (1,0.1);
\draw[dashed,lightgray] (0,0.2) -- (1,0.2);
\draw[dashed,lightgray] (0,0.3) -- (1,0.3);
\draw[dashed,lightgray] (0,0.4) -- (1,0.4);
\draw[dashed,lightgray] (0,0.5) -- (1,0.5);
\draw[dashed,lightgray] (0,0.6) -- (1,0.6);
\draw[dashed,lightgray] (0,0.7) -- (1,0.7);
\draw[dashed,lightgray] (0,0.8) -- (1,0.8);
\draw[dashed,lightgray] (0,0.9) -- (1,0.9);

\draw[dashed,lightgray] (0.1,0) -- (0.1,1);
\draw[dashed,lightgray] (0.2,0) -- (0.2,1);
\draw[dashed,lightgray] (0.3,0) -- (0.3,1);
\draw[dashed,lightgray] (0.4,0) -- (0.4,1);
\draw[dashed,lightgray] (0.5,0) -- (0.5,1);
\draw[dashed,lightgray] (0.6,0) -- (0.6,1);
\draw[dashed,lightgray] (0.7,0) -- (0.7,1);
\draw[dashed,lightgray] (0.8,0) -- (0.8,1);
\draw[dashed,lightgray] (0.9,0) -- (0.9,1);
   \draw[-,thick] (0,0) -- (1,0) node[right] {$q$};
   \draw[-,thick] (0,0) -- (0,1) node[above] {$p$};
   \draw[domain=0.5:1, range=0.01:1, smooth,variable=\x,red] plot ({\x},{ 2-1/\x });
      \draw[domain=0.2929:1, range=0.01:1, smooth,variable=\x,orange] plot ({\x},{ 2-1/(2*\x-\x*\x) });
      \draw[domain=0.2063:1, range=0.01:1, smooth,variable=\x,cyan] plot ({\x},{ 2-1/(3*\x-3*\x*\x+\x*\x*\x) });
     \draw[domain=0.159:1, range=0.01:1, smooth,variable=\x,green] plot ({\x},{ 2-1/(4*\x-6*\x*\x +4*\x*\x*\x - \x*\x*\x*\x) });   
     \draw[-,thick] (0,1)--(1,1);
     \draw[-,thick] (1,0)--(1,1); 
\end{tikzpicture}
\end{figure}
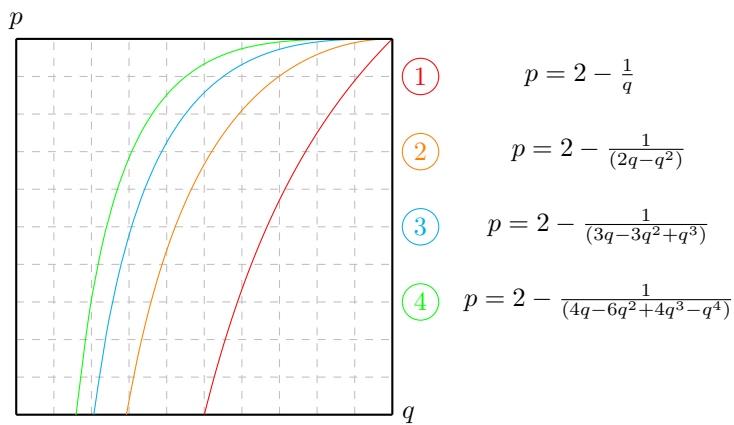

\noindent
Consider a slightly larger and denser network. 
\begin{figure}[ht!]
    \centering
      \caption{Village Graph 2}
\begin{tikzpicture}
\path
(0,0) node (1) [shape=circle,draw, inner sep=2,outer sep=.5] {1 }
(0,-.65) node (ip) [shape=circle,inner sep=.5,outer sep=0.5] {IP }
(2,1.5) node (2) [shape=circle,draw, inner sep=2,outer sep=.5] {2 }
(2,.5) node (3) [shape=circle,draw, inner sep=2,outer sep=.5] {3 }
(2,-.5) node (4) [shape=circle,draw, inner sep=2,outer sep=.5] {4 }
(2,-1.5) node (5) [shape=circle,draw, inner sep=2,outer sep=.5] {5 }
(5,1.5) node (6) [shape=circle,draw, inner sep=2,outer sep=.5] {6 }
(5,.5) node (7) [shape=circle,draw, inner sep=2,outer sep=.5] {7 }
(5,-.5) node (8) [shape=circle,draw, inner sep=2,outer sep=.5] {8 }
(5,-1.5) node (9) [shape=circle,draw, inner sep=2,outer sep=.5] {9 }
(7,0) node (10) [shape=circle,draw, inner sep=2,outer sep=.5] {10 }
;
\draw [dashed] (1)--(2);
\draw [dashed] (1)--(3);
\draw [dashed] (1)--(4);
\draw [dashed] (1)--(5);
\draw [dashed] (6)--(2);
\draw [dashed] (6)--(3);
\draw [dashed] (6)--(4);
\draw [dashed] (6)--(5);
\draw [dashed] (7)--(2);
\draw [dashed] (7)--(3);
\draw [dashed] (7)--(4);
\draw [dashed] (8)--(4);
\draw [dashed] (8)--(5);
\draw [dashed] (9)--(5);
\draw [dashed] (10)--(6);
\draw [dashed] (10)--(7);
\draw [dashed] (10)--(8);
\draw [dashed] (10)--(9);
\end{tikzpicture}
\end{figure}
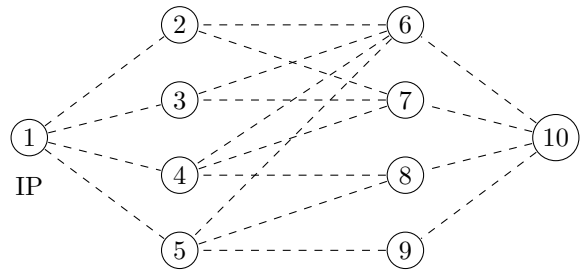
\noindent
What has changed from one graph to the other is a substantial increase in the number of intermediate PIIs. A large number of PIIs implies that the number of possible information propagation possibilities grows very large.
What can happen in any later information exchange always depends on what has happened in previous ones. 
Consequently, for each information scenario, we would, in each intermediate time period,  need to store the information who has been informed already in order to know who can be informed thereafter and to calculate the scenario specific individual information reception probabilities ($r_{it}$).
If we wanted to evaluate the exact log likelihood function that would imply storing $2^4$ period one information status vectors, for each of them computing and storing all possible period two information status vectors, in order to finally know - for each realization of $S_{1:2}$ - the final (scenario specific) information reception probabilities of all agents.
If the network is large and dense, it is our inability to store so many intermediate information status vectors and the time needed to process files that prevent us from establishing the log likelihood function.   
\\
\noindent 
If we have not eight, but roughly sixty to a hundred intermediate PIIs, then there would be not a billion or trillion, but a quintillion or nonillion of possible realizations of $S_{1:2}$. 
\\ \noindent 
To tackle the problem, we apply a trimming strategy. 
We will only allow a fixed number of $d$ PIIs to have two possible information status variables in the next period. For all other PIIs, the next period's information status will be set to a default (either one, in which case we assume the agent to be informed and opted out, or zero in which case we presume her to be uninformed). In particular, the individual will be trimmed to whichever of the two scenarios is more likely in the given period. 
If in each of the first two exchanges, the maximal number of PIIs is set to a trimming value of $d$, then we end up with a number of $s_{1:2}$ scenarios that cannot exceed $2^{2d}$. Storing information scenarios is only necessary to calculate the probabilities of what can follow. Since nothing follows after the third information exchange, a large number of final PIIs is absolutely un-problematic and there is no need to restrict the last wave of PIIs. 
 \\
\noindent Accordingly, we need to select $d$ PIIs for which the two scenarios will be considered. Recapitulating that each PII can, in each exchange in which she can reached be either newly informed and uninterested or uninformed, we can see that there exists a threshold $r_{it}$ such that both are equally likely.
\[ \underbrace{(1-p) r_i}_{
\substack{
\mbox{probability of scenario A:}\\
\mbox{$i$ is informed and opts out}
}
} = 
\underbrace{(1-r_i)}_{
\substack{
\mbox{probability of scenario B:}\\
\mbox{$i$ is uninformed}
}
} \]
\begin{equation}
\Rightarrow 
2- \frac{1}{r_i} = p
\end{equation}
The simple trimming rule identifies agents that are the furthest away from this threshold and trims their information status to either zero (if the agent's information reception probability falls below the threshold) or one (if it exceeds the threshold). This is repeated until only $d$ PIIs are left unrestricted. The basic idea of the proceeding is that if an agent is far away from the threshold, the probability of one scenario will be large while the other will be small and as such, neglecting the less likely scenario for these agents is the optimal choice from the current perspective (i.e.\ not taking into account what can happen thereafter) in the sense that the currently most likely scenarios are chosen. It may be the case, however, that a trimming choice reveals itself as sub-optimal in one of the following periods. Most dramatically, imagine a period-3 participant that is two links away from the nearest IP. 
The algorithm may set {\it} all neighbors of her neighbors to the default status “uninformed" in the first information exchange, which obviously has probability zero given that {\it someone} must have informed the period-3 participant. In the Appendix, we discuss under which circumstances such erroneous choices occur and how they depend on the network topology.   
\\
\noindent 
We demonstrate the trimming with the second graph above. Assume we are at a grid-point $p,q$ for which $2-\frac{1}{q}>p$ implying that scenario A is more likely for any PII and that we pick a $d$ value of 2. All PIIs featuring the same number of IP links in the first information exchange, the algorithm arbitrarily picks individual 2 and 3 to be trimmed. With 2 and 3 being newly informed and opting out, there are 4 scenarios resulting from the first exchange. Depending on which scenario was realized, there are four or five period-two PIIs. In any case, however, 6 and 7 will always be the ones with the largest number of links to the information and hence the algorithm trims them to scenario A. In the last exchange, there will be three or four PIIs. No individual will have more links to the information than 10 and as such, she will be subject to trimming. If there are four PIIs, then another PII to be trimmed will be chosen from the remaining three ones.  
\\ \noindent Assume next that the grid-point is such that $2-\frac{1}{(1-(1-q)^4)} < p$ implying that scenario B is more likely for any PIIs and that we again pick a $d$ value of 2. This time, 2 and 3 will be trimmed to being uninformed in the first exchange. As such, the second exchange features four to six PIIs. No PII can have less links to the information than 2 and 3 and thus they will be trimmed again and not receive the information. Whenever there are six PIIs, two additional PIIs that have one link to the information will be trimmed. The last exchange again trimms 2 and 3 and if necessary up to three other PIIs.\footnote{Trimming an individual to zero implies that she temporarily looses her right to receive the information or that the link is temporarily deactivated. She can, however, enter the PIIs in the next period. Whether or not she will be cut again depends on her and everybody else's probability to receive the information in that period. }
\\ \noindent For each link portfolio an agent can have, there exist a specific $pq$ relationship such that the two scenarios are equally likely. Figure \ref{thresh} illustrates this with a link portfolio of one, two, three and four links.  
\noindent 
The trimming strategy works better when agents are furthest away from the threshold rate.  
As a direct consequence, the best performance can be expected whenever either $p$ or $q$ is small and the respective other large. 
  \\
\noindent 
Worth mentioning, note that the number of scenarios considered is endogenous. Each of the $2^d$ period-1 information status vectors entails a different number of scenarios that are subsequently possible as the umber of agents in reach of the information and the number of agents already informed vary. As a consequence, each trimming choice induces a different number of scenarios taken into account when establishing the log likelihood function.   
The rates at which trimming occurs are endogenous too. If the aim is to always trim to a certain number of PIIs, thus the lower and upper trimming rates vary on the grid.   

\section{Monte Carlo Set-up}

The aim of the study is to evaluate the performance of the simple trimming technique as the trimming value ($d$) varies.

\begin{table}[ht!]
    \centering
        \caption{Monte Carlo Parameters}
    \begin{tabular}{c|c}
    \hline 
       $N$  & size of the networks: 20 \\
       seed S  & seed for drawing the IP \\
       & ranging from 1 to 90 \\
       seed D & seeds for simulating the data \\
       & ranging from 2 to 91 \\
       $p_0$ & participation rate used to simulate the data \\
       $q_0$ & diffusion rate used to simulate the data \\
       $V$ & sample size: 11 \\
       $R$ & number of replications: 90 \\
    \end{tabular}
    \label{tab:my_label}
\end{table}
\noindent
We use eleven network adjacency matrices from \citet{BCDJ}. These were surveyed by asking villagers in India with whom they engage in regular activities. We here use the union of the activity-specific networks, thus assuming that any joint activity indicates the presence of a link.
The alternative would be to simulate the network data as well. However, this would require us to specify a network formation model and to choose its parameters and these choices would impact the study results.   \\
\noindent The eleven villages are large and dense. In the Monte Carlo study, we need to be able to evaluate the exact log likelihood function. To achieve tractability, we pick a sub-matrix consisting of $N$ individuals. By the same token, to limit the number of PIIs, only one individual will obtain the information initially. This individual is randomly drawn from the $N$ villagers. The sub-matrix used in each replication is different.\footnote{The algorithm starts at the element row seed S, column seed S situated on the diagonal and reads out a square 20 times 20 matrix}
\\ \noindent
Each repetition consists of 
\underline{simulation} and \underline{estimation}.
While $N$, $p_0$, $q_0$ and $V$ are held fixed, each repetition distinguishes itself by its unique combination of the seeds used to initialize the random number generators. We consider the settings: case 1: $p_0=0.5, q_0=0.5$, case 2: $p_0=0.1, q_0=0.9$ and case 3: $p_0=0.9, q_0=0.1$ with $N=20$ and $V=11$. \\
\\
\noindent 
\underline{Simulation:} 
The participation data is simulated according to the model assumptions using parameters $p_0$ (participation probability) and $q_0$ (information transmission probability) and the network sub-matrices. Each replication chooses a different sub-matrix out of the network matrix. Following the random draw of the initially informed individual and the random determination of her participation status, there are three rounds of information diffusion and participation. Each round, 
for each individual linked to the information, we draw a number from the uniform distribution in the interval $[0,1]$ and specify that the information is passed on if the number drawn is smaller or equal to $q_0$. Equivalently, for each newly informed individual, the number drawn from the standard uniform distribution needs to be smaller or equal to $p_0$ for her to participate. \\
\noindent
This leads to a participation data matrix for each village, i.e.\ one sample. \\
\\
\noindent 
\underline{Estimation:}
The log likelihood function is established as the sum over information scenarios and each scenarios has a different number of potentially informed individuals (PIIs). Our trimming strategy sets a maximum to this number and sets PIIs exceeding this number to their more likely scenario in the first two exchanges (trimming is computationally  unnecessary in the last exchange). In the simulation study, we wish to let the trimming value ($d$) vary and evaluate the consequences. Therefore, 
for each village, we find the maximal number of PIIs ($\bar{d}_v$). We then compute the approximate log likelihood function for $d=0,...,\bar{d}_v-1$ (denoted $\mathcal{L}_{v,d}$) and the correct log likelihood function, i.e.\ $d=\bar{d}_v$ which implies no trimming (denoted $\mathcal{L}_{v,\bar{d}_v }$). Each time, the number of PIIs is trimmed down to size $d$ in the first two information exchanges. Then the approximate sample log likelihood function for each trimming value is computed by aggregating all villages, i.e. 
\[ \mathcal{L}_d= \sum_{v=1}^V \mathcal{L}_{v,max(d, \bar{d}_v)} \]
(i.e. if the trimming value exceeds the maximal number of PIIs in the village, the village is not subject to trimming). 
Changes of the trimming value when it is low hence impact practically all villages and in the villages practically all scenarios but as the trimming value grows towards its maximum, only few villages and scenarios are actually subject to trimming. 
Then the peak of the log likelihood function for each $d$ is found by grid search and confidence sets are established using the LR test. Each repetition hence delivers a sequence of estimates $\hat{p}_d,\hat{q}_d$ together with their confidence sets for $d=0,...,max(\bar{d}_v)$. We compare the estimates at different trimming values with the estimates that result from applying ML to a shortened time horizon of two periods.  

\section{Monte Carlo Results}

\begin{figure}[H]
     \centering
       \caption{Case 1: $p_0=0.5, q_0=0.5$ Estimates and Standard Deviations}
       \begin{subfigure}{0.45\textwidth}
       \includegraphics[width=\textwidth]{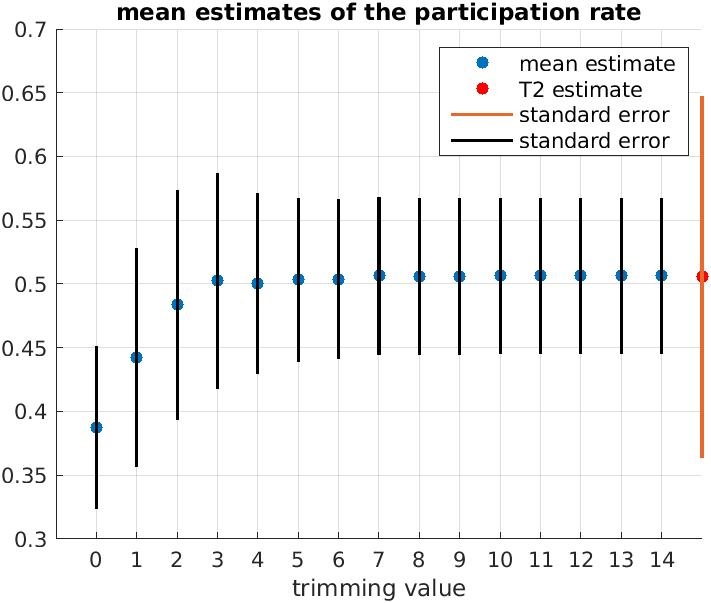}
       \end{subfigure}
             \begin{subfigure}{0.45\textwidth}
       \includegraphics[width=\textwidth]{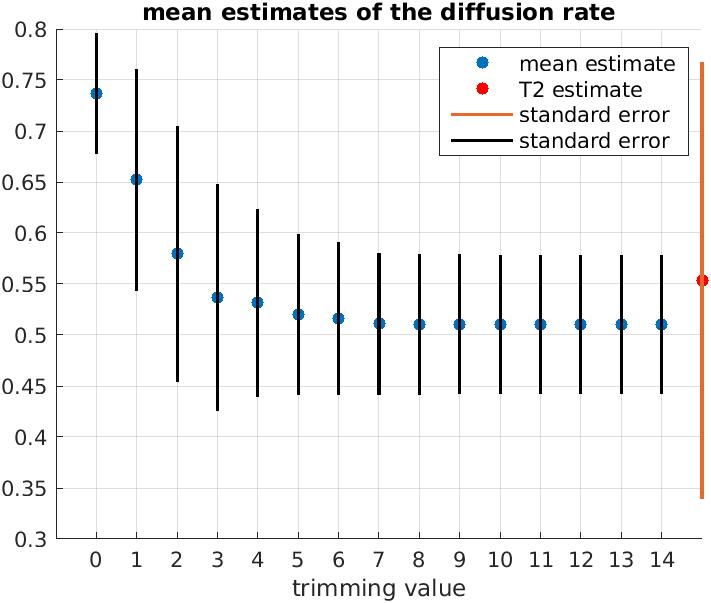}
       \end{subfigure} 
    \label{est0505}
 \end{figure}

\begin{figure}[H]
     \centering
       \caption{Case 1: $p_0=0.5, q_0=0.5$,  }
       \begin{subfigure}{\textwidth}
       \caption{Estimated Densities $p$ for Different Trimming Values (black) and the Two-period MLE (orange)}
            \includegraphics[width=\textwidth]{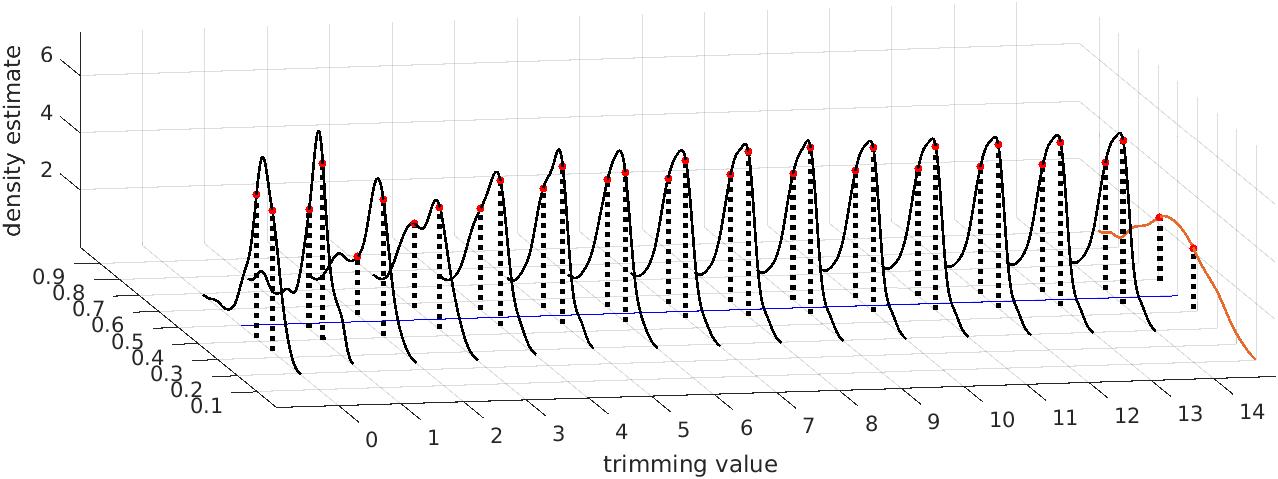}
       \end{subfigure} \\
         \begin{subfigure}{\textwidth}
          \caption{Estimated Densities $q$ for Different Trimming Values (black) and the Two-period MLE (orange)}
            \includegraphics[width=\textwidth]{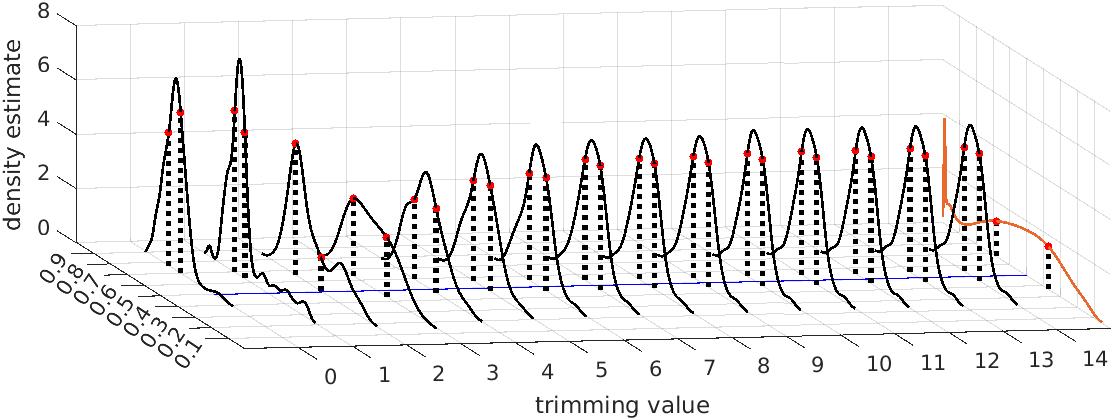}
       \end{subfigure}
     \label{dens0505}
 \end{figure}

\begin{figure}[H]
     \centering
       \caption{Case 1: $p_0=0.5, q_0=0.5$ Histograms (Trimming Estimator)}
       \begin{subfigure}{\textwidth}
       \caption{Histograms $p$}
       \includegraphics[width=\textwidth]{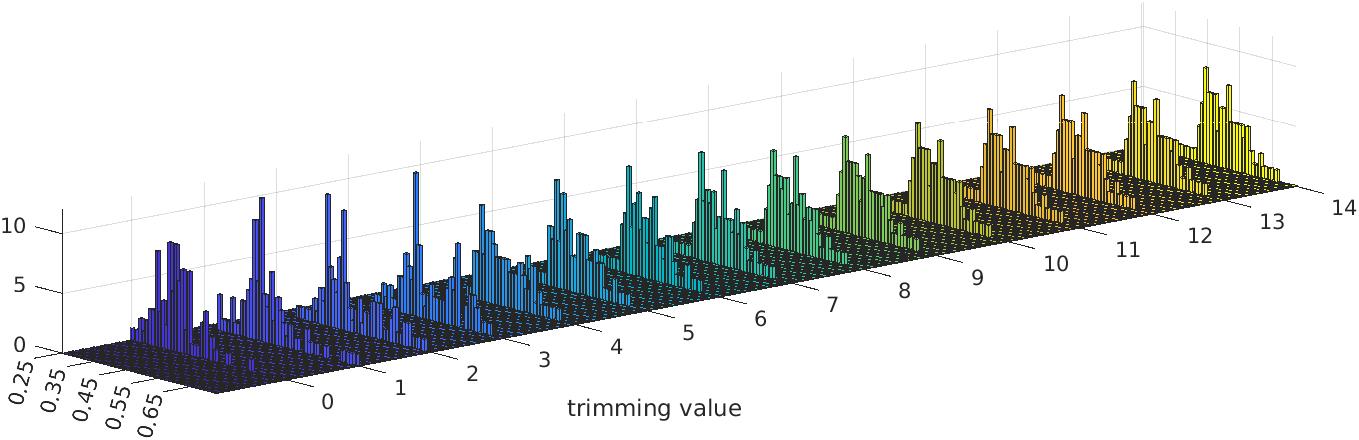}
       \end{subfigure}\\
            \begin{subfigure}{\textwidth}
            \caption{Histograms $q$}
       \includegraphics[width=\textwidth]{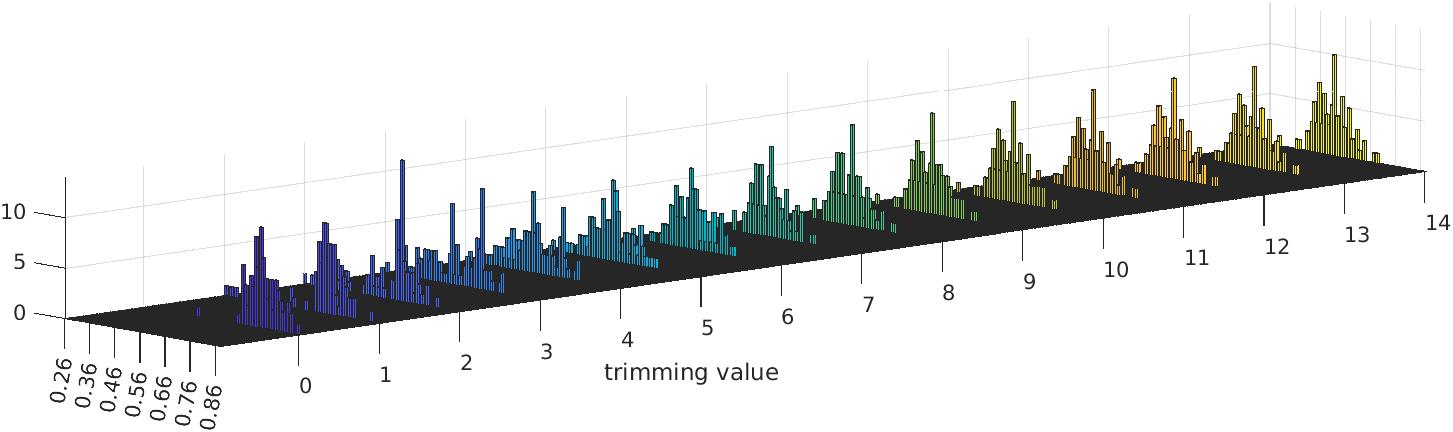}
       \end{subfigure}
    \label{hist0505}
 \end{figure}

\begin{figure}[H]
     \centering
       \caption{Case 1: $p_0=0.5, q_0=0.5$ Histograms (Two-period MLE) }
       \begin{subfigure}{0.45\textwidth}
       \includegraphics[width=\textwidth]{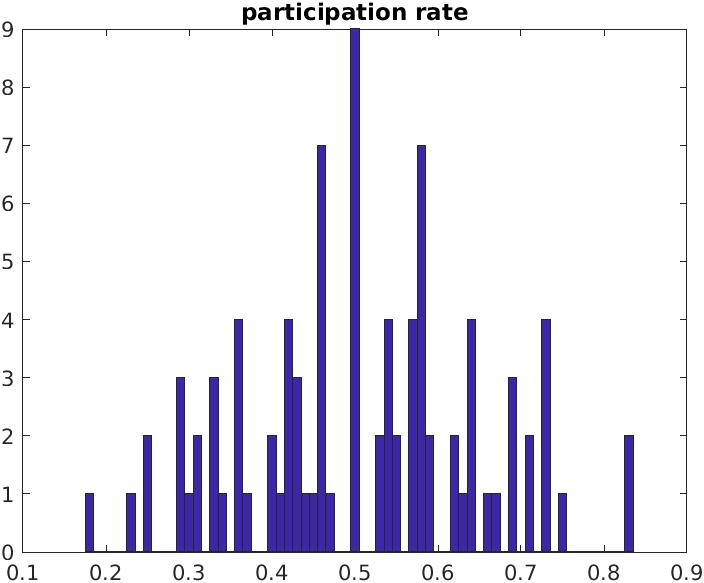}
       \end{subfigure}
        \begin{subfigure}{0.45\textwidth}
       \includegraphics[width=\textwidth]{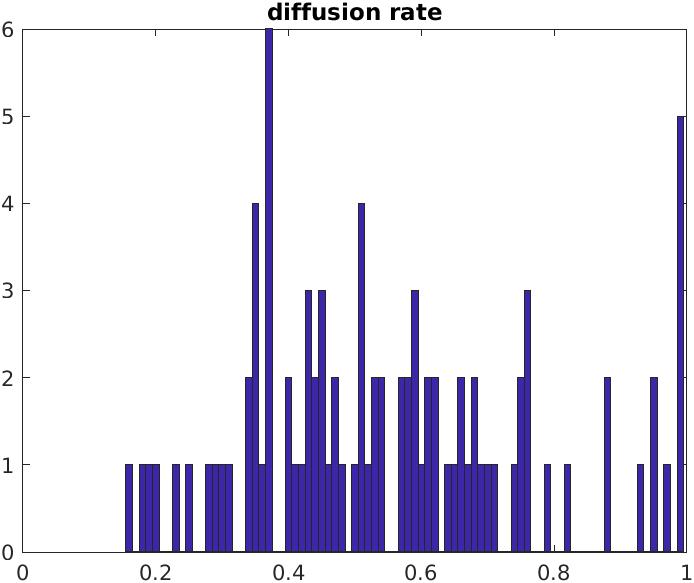}
       \end{subfigure}
     \label{t20505}
 \end{figure}

\begin{figure}[H]
     \centering
       \caption{Case 2: $p_0=0.1, q_0=0.9$ Estimates and Standard Deviations}
       \begin{subfigure}{0.45\textwidth}
       \includegraphics[width=\textwidth]{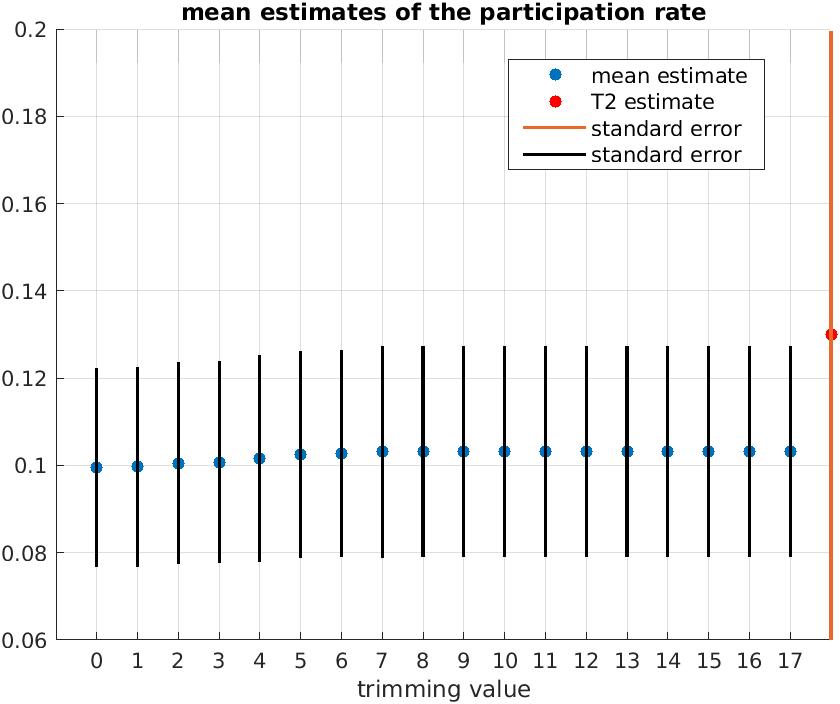}
       \end{subfigure}
             \begin{subfigure}{0.45\textwidth}
       \includegraphics[width=\textwidth]{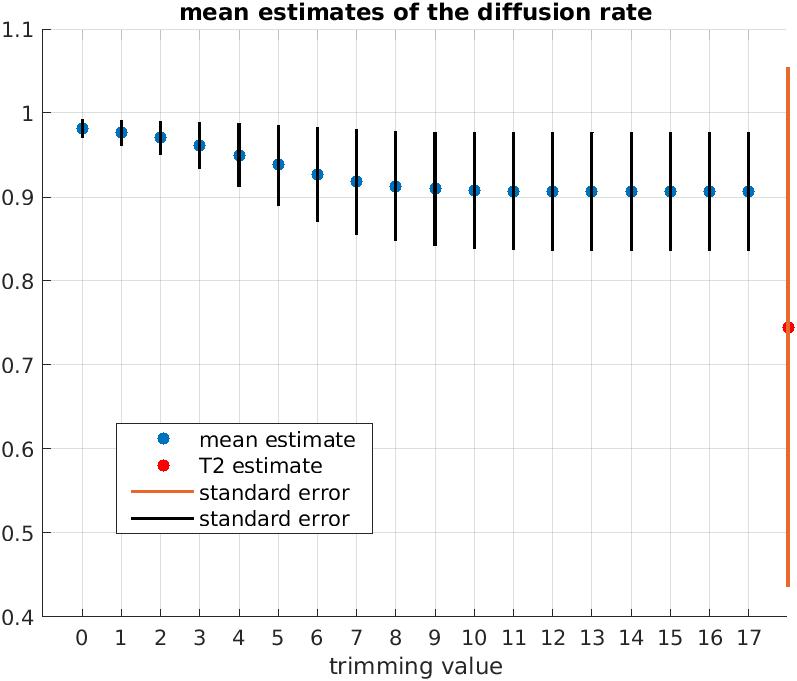}
       \end{subfigure} 
     \label{est0109}
 \end{figure}

\begin{figure}[H]
     \centering
       \caption{Case 2: $p_0=0.1, q_0=0.9$,  }
       \begin{subfigure}{\textwidth}
       \caption{Estimated Densities $p$ for Different Trimming Values (black) and the Two-period MLE (orange)}
            \includegraphics[width=\textwidth]{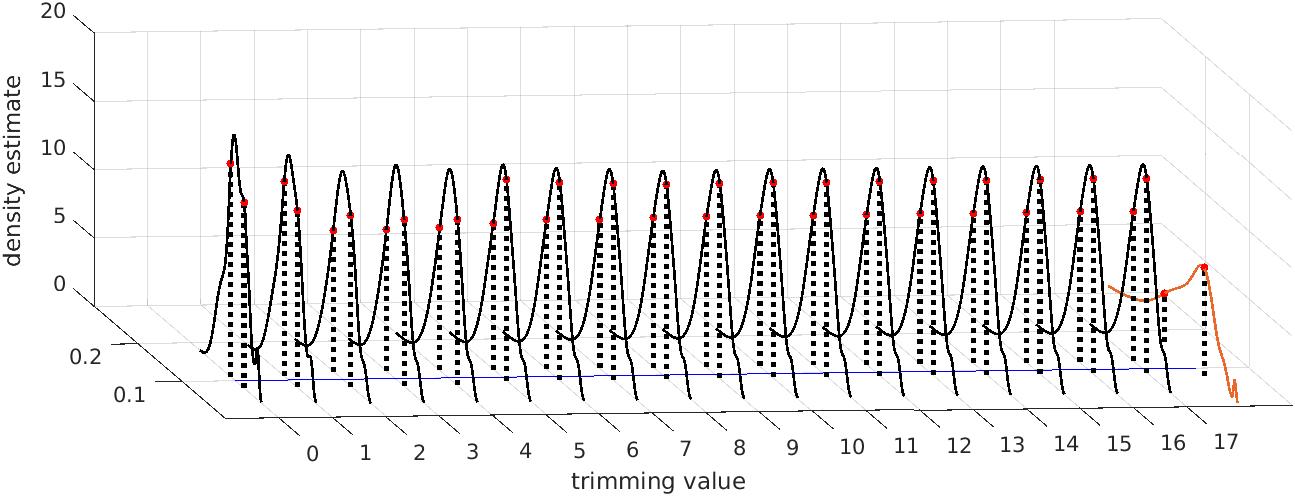}
       \end{subfigure} \\
         \begin{subfigure}{\textwidth}
          \caption{Estimated Densities $q$ for Different Trimming Values (black) and the Two-period MLE (orange)}
            \includegraphics[width=\textwidth]{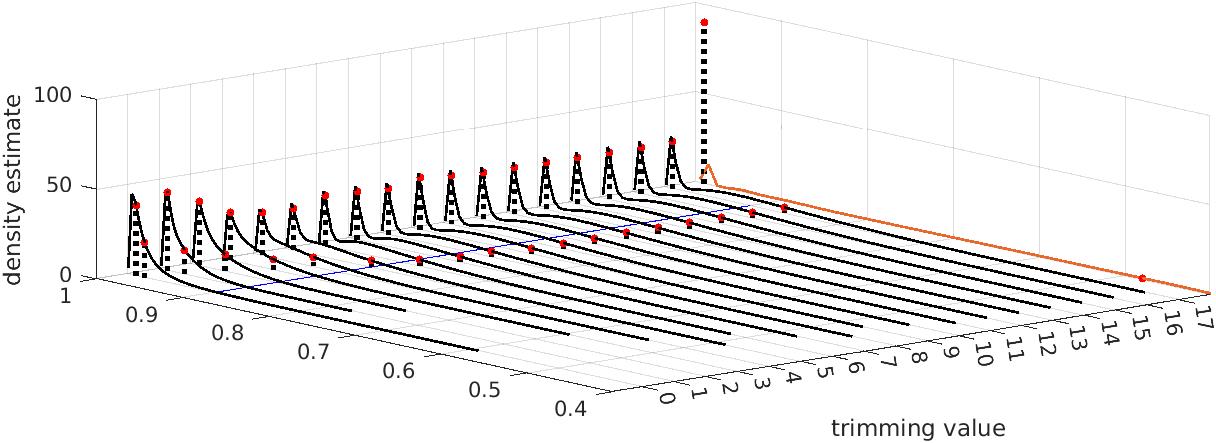}
       \end{subfigure}
    \label{dens0109}
 \end{figure}

\begin{figure}[H]
     \centering
       \caption{Case 2: $p_0=0.1, q_0=0.9$ Histograms (Trimming Estimator)}
       \begin{subfigure}{\textwidth}
       \caption{Histograms $p$}
       \includegraphics[width=\textwidth]{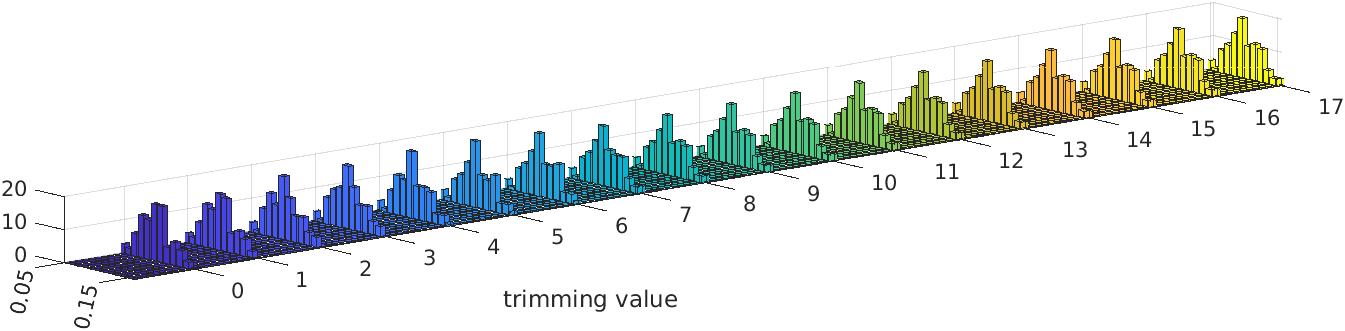}
       \end{subfigure}\\
            \begin{subfigure}{\textwidth}
            \caption{Histograms $q$}
       \includegraphics[width=\textwidth]{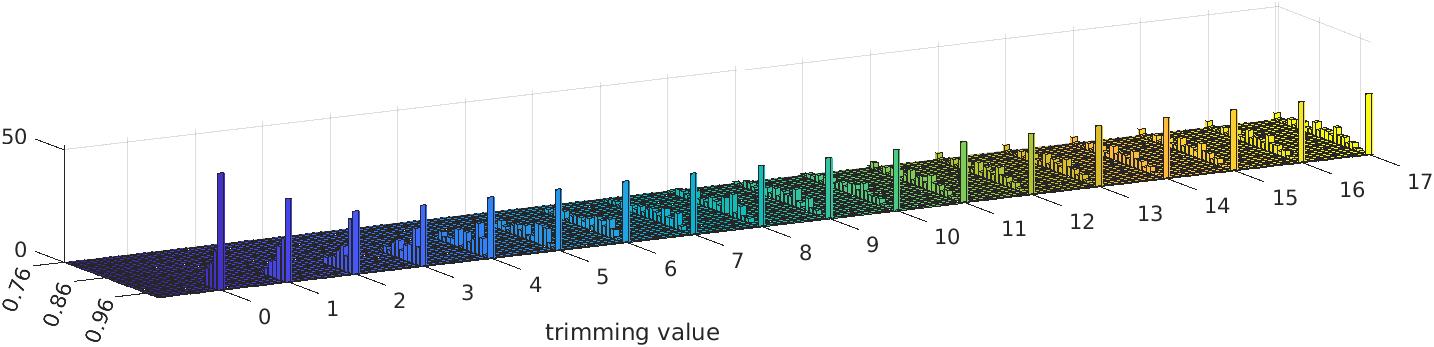}
       \end{subfigure}
     \label{hist0109}
 \end{figure}

\begin{figure}[H]
     \centering
       \caption{Case 2: $p_0=0.1, q_0=0.9$ Histograms (Two-period MLE) }
       \begin{subfigure}{0.45\textwidth}
       \includegraphics[width=\textwidth]{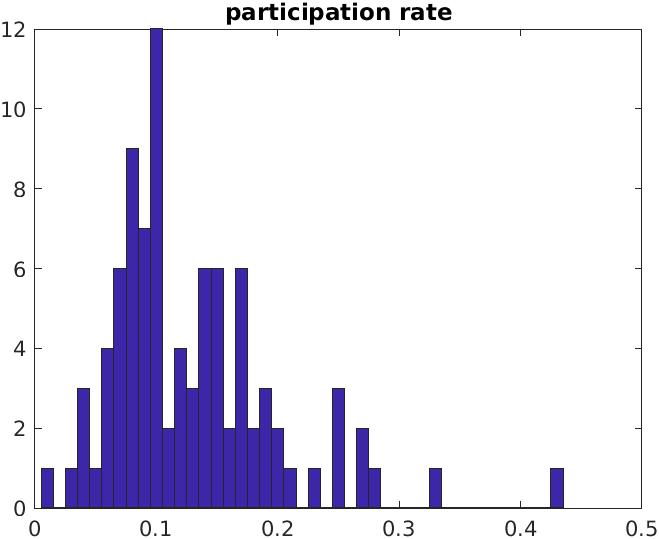}
       \end{subfigure}
        \begin{subfigure}{0.45\textwidth}
       \includegraphics[width=\textwidth]{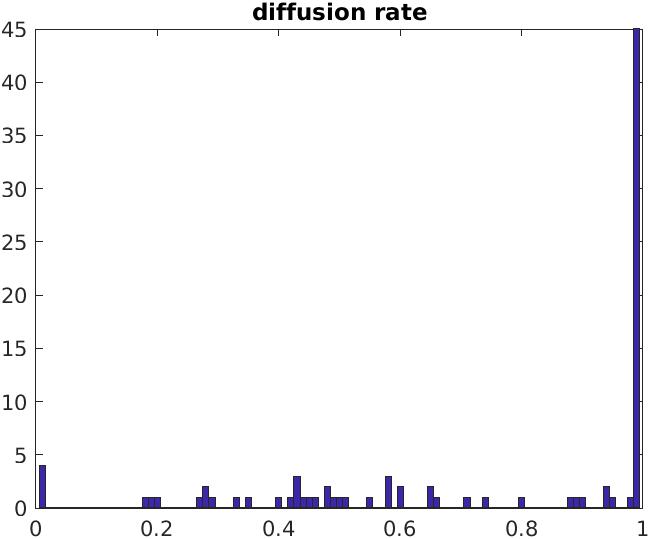}
       \end{subfigure}
     \label{t20109}
 \end{figure}

\begin{figure}[H]
     \centering
       \caption{Case 3: $p_0=0.9, q_0=0.1$ Estimates and Standard Deviations}
       \begin{subfigure}{0.45\textwidth}
       \includegraphics[width=\textwidth]{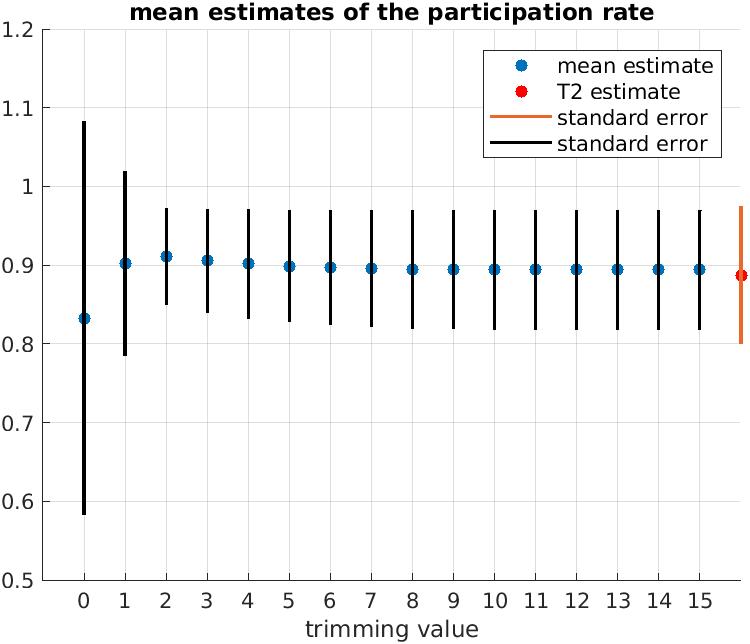}
       \end{subfigure}
             \begin{subfigure}{0.45\textwidth}
       \includegraphics[width=\textwidth]{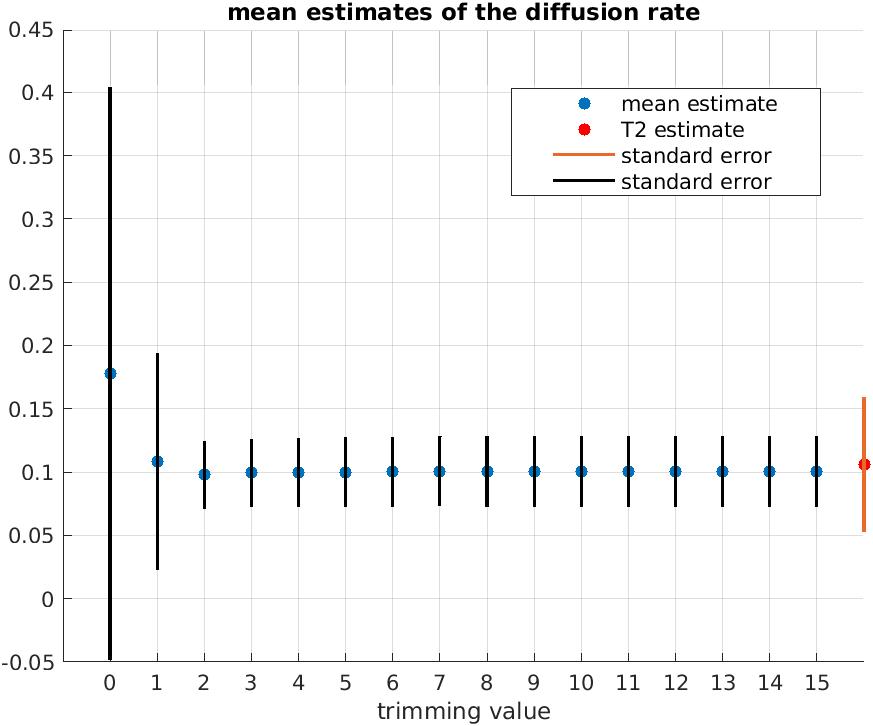}
       \end{subfigure} 
     \label{est0901}
 \end{figure}

\begin{figure}[H]
     \centering
       \caption{Case 3: $p_0=0.9, q_0=0.1$,  }
       \begin{subfigure}{\textwidth}
       \caption{Estimated Densities $p$ for Different Trimming Values (black) and the Two-period MLE (orange)}
            \includegraphics[width=\textwidth]{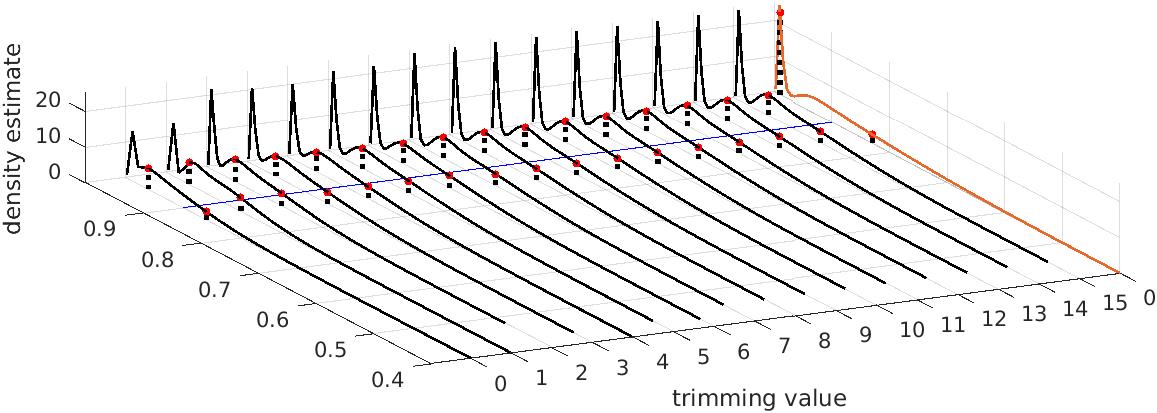}
       \end{subfigure} \\
         \begin{subfigure}{\textwidth}
          \caption{Estimated Densities $q$ for Different Trimming Values (black) and the Two-period MLE (orange)}
            \includegraphics[width=\textwidth]{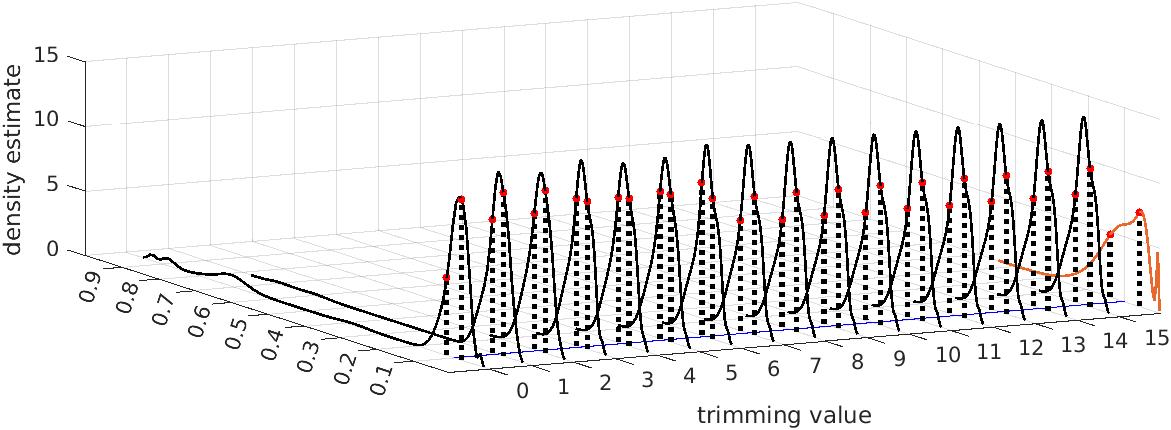}
       \end{subfigure}
    \label{dens0901}
 \end{figure}

\begin{figure}[H]
     \centering
       \caption{Case 3: $p_0=0.9, q_0=0.1$ Histograms (Trimming Estimator)}
       \begin{subfigure}{\textwidth}
       \caption{Histograms $p$}
       \includegraphics[width=\textwidth]{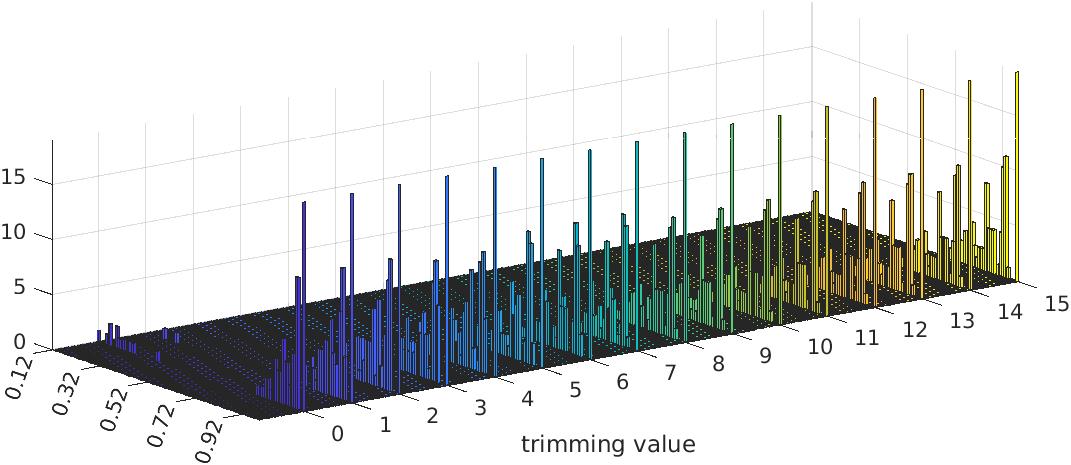}
       \end{subfigure}\\
            \begin{subfigure}{\textwidth}
            \caption{Histograms $q$}
       \includegraphics[width=\textwidth]{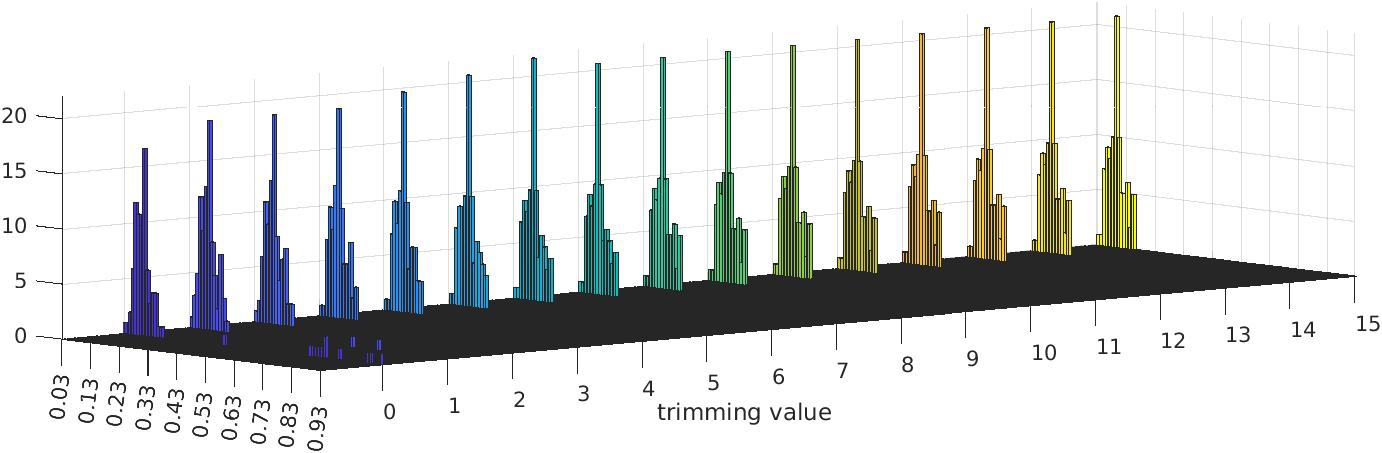}
       \end{subfigure}
    \label{hist0901}
 \end{figure}

\begin{figure}[H]
     \centering
       \caption{Case 3: $p_0=0.9, q_0=0.1$ Histograms (Two-period MLE) }
       \begin{subfigure}{0.45\textwidth}
       \includegraphics[width=\textwidth]{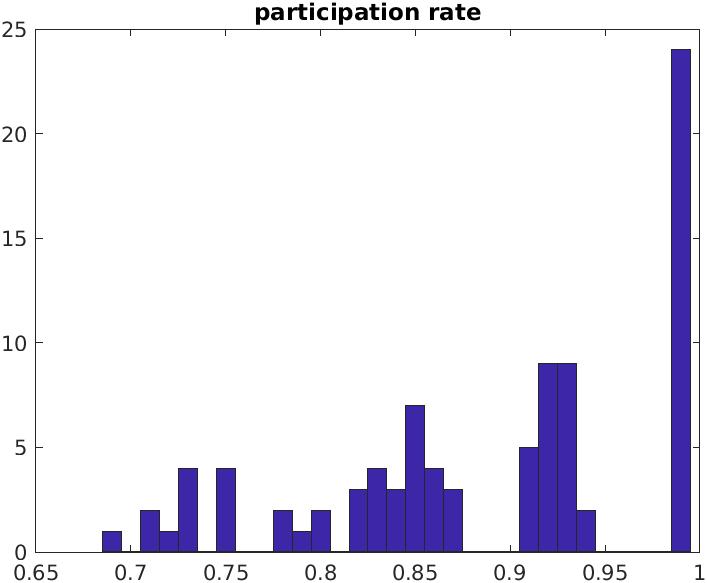}
       \end{subfigure}
        \begin{subfigure}{0.45\textwidth}
       \includegraphics[width=\textwidth]{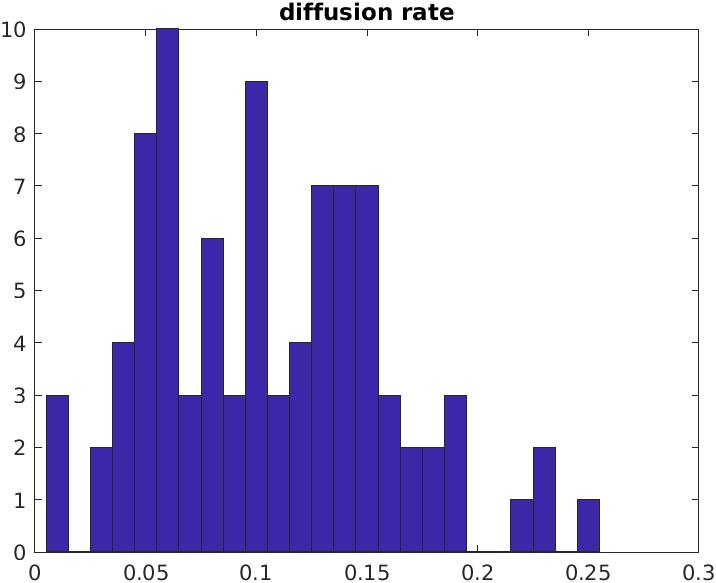}
       \end{subfigure}
     \label{t20901}
 \end{figure}

For each case, the proceeding described above leads to ninety series of estimates. Remembering that trimming takes place whenever the number of PIIs exceed the desired trimming value, thus the lengths of this series is pinned down by the largest number of PIIs in any village and the last estimate of this series is the four-period MLE. Additionally, we dispose over the MLE that results from restricting the time horizon to two periods (i.e. the “Two-period Estimator" from chapter two).\\ \noindent  
In Figures \ref{est0505}-\ref{t20901}, we plot the mean estimates and their empirical standard errors, the estimated density and the histograms of the empirical probability mass function. 
\\ 
\noindent 
For each case the first two plots depicts the mean estimate ($\Bar{\hat{p}}$ and $\Bar{\hat{q}}$) as a function of the trimming value, the mean being taken over the 90 estimates (Figures \ref{est0505}, \ref{est0109} and \ref{est0901}). The vertical bars have the length of the empirical standard error in each direction, the latter being simply computed as the square root of average squared deviation of the individual estimate from the mean estimate. 
\[ \left( \frac{1}{89}
\sum_{r=1}^{90}
(\hat{p}_r-
\Bar{\hat{p}})^2 \right)^{1/2} 
\hspace{2cm}
 \left( \frac{1}{89}
\sum_{r=1}^{90}
(\hat{q}_r-
\Bar{\hat{q}})^2 \right)^{1/2}
\]
The trimmed estimates converge to the MLE rather quickly. However, it needs to be kept in mind that we have restricted the village sizes to twenty and as a result, a substantial fraction of villages simply feature very few PIIs, in particular when participation rates are not too small.  
Then, these villages are already at the MLE, even for minor trimming values. The results are nonetheless encouraging in the sense that trimming substantially increases computational speed and does not hurt much (compared to the MLE) as long as the sample includes enough villages for which no more than one third of the PIIs are trimmed. The efficiency gains compared to the “Two-period Estimator" are substantial, in particular for the diffusion rate.  \\
 \noindent
Thereafter, for each case, we depict depict the kernel density estimates together with the empirical quantiles (first and third)(Figures \ref{dens0505}, \ref{dens0109} and \ref{dens0901}). The density of the “Two-period Estimator", depicted in orange, is substantially flatter and more often exhibit multiple local maxima. \\
 \noindent 
The histograms in the third plot series (Figures \ref{hist0505}, \ref{hist0109}, \ref{hist0901} and \ref{t20505}, \ref{t20109}, \ref{t20901}) reconfirm that the “Two-period estimates" are much more dispersed, especially for the diffusion rates. As the Trimming estimates converge, the histograms for larger trimming values become indistinguishable from the MLE.  
 \\ \\
 \noindent 
 The following case-specific observations can be made. \\ 
 {\bf Case 1:} \\
\noindent 
The densities for “Trimming estimates" exhibit a distinct peak in the probability mass function, which is not the case for the “Two-period estimates". Observe that for the latter, the estimated diffusion rate for a substantial fraction of the sample is at 0.99. Also, the estimates are much less dispersed for the “Trimming-Estimator" than the “Two-period Estimator".  
 \\
 \noindent
{\bf Case 2:}  \\
 \noindent
 The gains from increasing the trimming value are made exclusively for the diffusion rate estimates, while the participation rate estimates are very similar.  \\
\noindent The standard deviations are smaller and the log likelihood surfaces are more peaked for small trimming values. 
\\
\noindent
{\bf Case 3:} \\
Both $p$ and $q$ are well identified whenever adoption rate is  high (except for the extreme case of $d=0$, which would imply trimming all PIIs). Unsurprisingly, in this case also the performance difference as compared to the “Two-Period estimator" is smallest.

\section{Real Data Application}
This section presents the results of an application of the “Trimming-estimator" to four selected real and full-size villages. This estimation was run on the powerful Amazon-Web-Service server cluster and I am  grateful that funds have been made available for this estimation, which also provided me with the opportunity to gain valuable experience with respect to the usage of this server cluster.\\ \noindent 
The time horizon is set to 3, the chosen villages are 1, 12, 31 and 67. The choice was guided by the fact that in particular these four villages exhibit a steady increase in the observed real adoption rates, whereas for many other villages there is a distinct jump. With jumps being potentially associated to measurement error, I hope to have chosen the villages with the most informative observed adoption patterns. These villages have, respectively, 182, 175, 153 and 193 inhabitants. \\ \noindent
Due to the computational cost, it was infeasible to calculate the objective function over the entire grid. However, this was also not necessary. Extensive testing indicated that the convergence was relatively slow (approximately 0.01 when the trimming value increases by two for $q$ and practically nonexistent for $p$). A relatively low trimming value of eighteen was chosen to define the subgrid to be evaluated. The subgrid contains 48 point with the diffusion rate varying from 0.58 to 0.73 and the adoption rate from 0.23 to 0.25. This range was chosen according to a prognostic calculation such that the resulting grid would cover approximately most of the five present confidence set of the final largest feasible trimming value, which was 32. \\ \noindent
For a trimming estimation with three time periods that starts from a trimming value that is low compared to the total number of villagers, the computation time roughly doubles when the trimming value is increased by one. Parallelization was done over both villages and gridpoints. The final computation times were 651.2 hours (village 1), 609.5 hours (village 12), 480.7 hours (village 31) and 751.8 hours (village 67).

\begin{table}[H]
\caption{Sequence of Trimming Estimates}
    \centering
    \begin{tabular}{c|c|c|c|c|c|c}
         &d=18&d=28&d=29&d=30&d=31&d=32  \\ \hline
$p$     & 0.24&	0.25&	0.25&	0.25&	0.25&	0.25\\ \hline
$q$     &0.73&	0.68&	0.68&	0.67&	0.67&	0.66 \\\hline
5\% conf.&&&&&&\\
set $p$ &
$[0.23,0.25]$	& $[0.23,0.25]$ &	$[0.23,0.25]$ &	$[0.23,0.25]$ &	$[0.23,0.25]$ &	$[0.23,0.25]$ \\ \hline
5\% conf.&&&&&&\\ 
set $q$ &
$[0.66,0.73]$&	$[0.61,0.73]$&	$[0.6,0.73]$&	$[0.6,0.73]$&
$[0.59,0.73]$& $[0.59,0.73]$\\
    \end{tabular}
    \label{tab:my_label}
\end{table}
\noindent
The fact that the estimate is situated at the border of the grid is not problematic given that testing indicated that convergence of $p$  was achieved much faster than convergence of $q$ and as such, not much change can be expected in the adoption parameter when the trimming value further increases. The confidence sets are enlarging when the trimming value increases. This is intuitive: the subgrid around the peak is situated in an area in which all agents are trimmed to being “informed and opted out” (the small $p$, high  $q$ area). A change in $q$ will thus have the same effect on all trimmed individuals leading to a more pronounced peak, while in the full likelihood, the impact of a change in $q$  would be different for different scenarios, thus resulting overall in a flatter log likelihood function. This hints towards the finding that indeed the trimming estimate may, in cases of relatively low trimming values, lead to an over-pronounced peak such that it may be necessary to adjust critical values accordingly. This same finding is also in the contour and surface plots. \\ \noindent 
The “Two-period Estimate" estimate is outside the interval considered. Given the test results, it is expected that the full MLE with three time periods will converge to a value that is somewhere in between the trimming and the “Two-period estimate". \\ \noindent 
Estimation using only four villages is highly challenging as we can deduct from the sizes of the “Two-period estimate"’s confidence sets. Note that when testing on the basis of the “Two-period estimate", the hypothesis that it equals the trimming estimate cannot be rejected even at 30\%.
\\ \noindent
Since practically all points in the sub-grid are in the 5\% confidence set, hence I compute the confidence sets for higher confidence values. As compared to the “Two-period  estimate", the “Trimming estimate"’s confidence set is extremely narrow. This is in line with the observation that the confidence sets widen when the trimming value is increased.

\begin{figure}[H]
   \caption{Confidence Sets Based on the Two-Period estimate of Chapter 2}
    \includegraphics[scale=0.25]{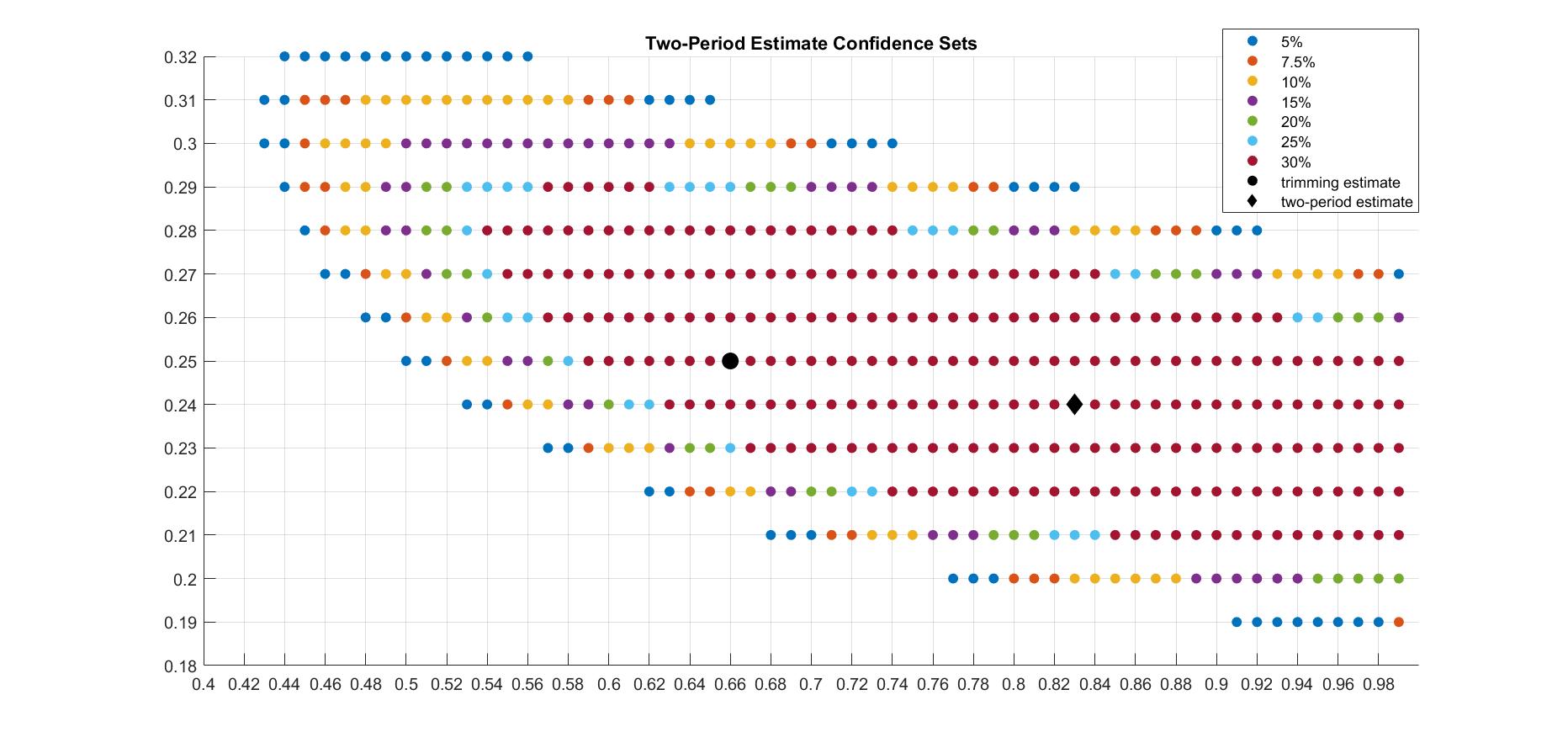}
    \label{fig:my_label}
\end{figure}

\begin{figure}[H]
     \caption{Confidence Sets Based on the Trimming Estimate}
    \includegraphics[scale=0.25]{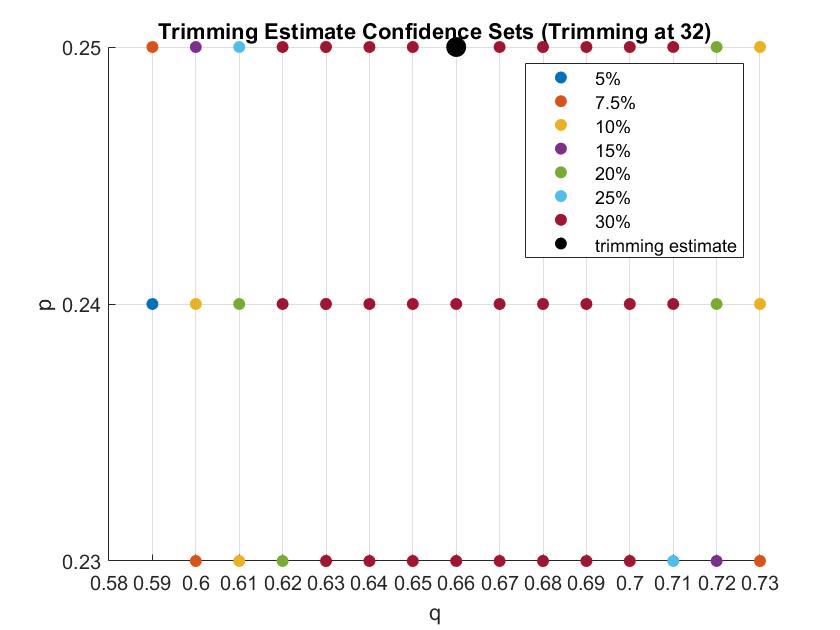}
    \includegraphics[scale=0.25]{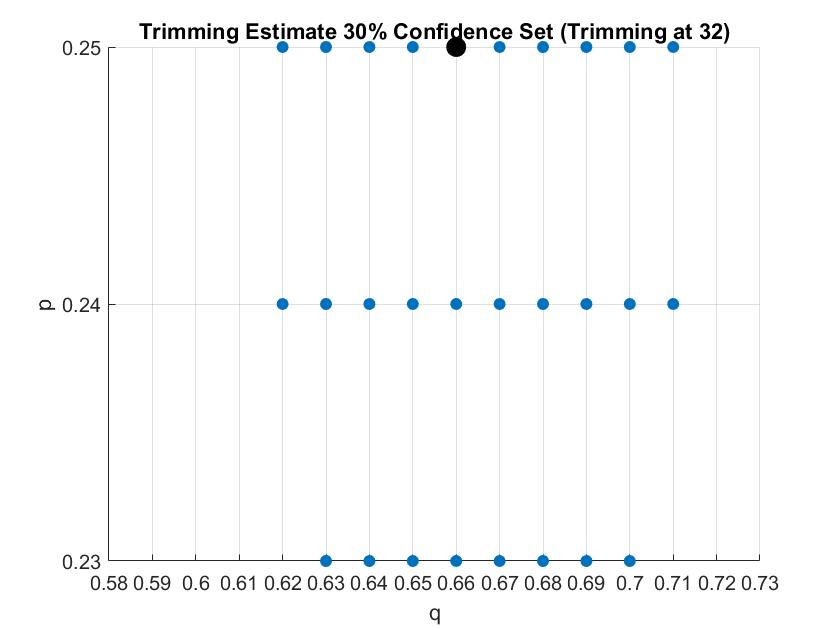}
    \label{fig:my_label}
\end{figure}

\begin{figure}[H]
     \caption{Likelihood Surface and Contour Based on the Trimming Estimate}
    \includegraphics[scale=0.25]{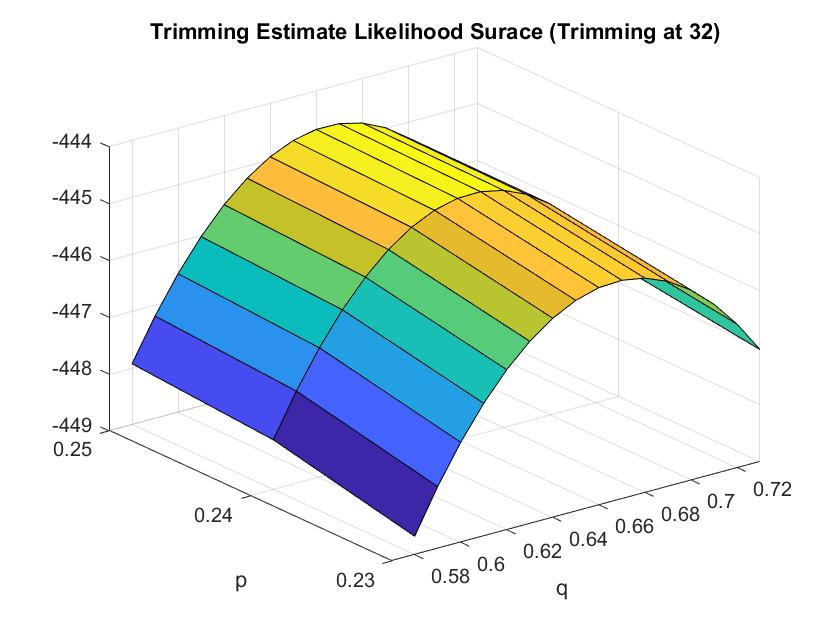}
     \includegraphics[scale=0.25]{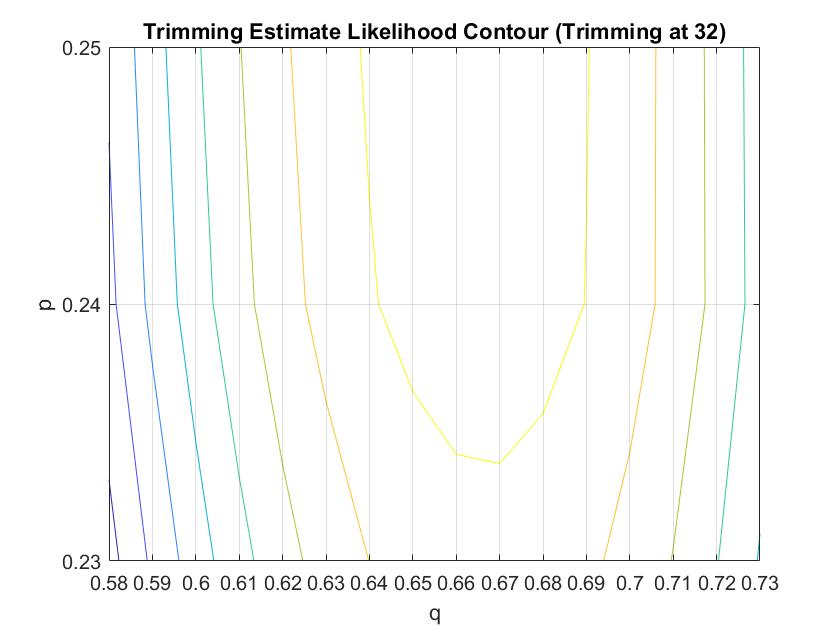}
    \label{fig:my_label}
\end{figure}

\begin{figure}[H]
     \caption{Likelihood Surface and Contour Based on the Two-Period Estimate from Chapter 2}
    \includegraphics[scale=0.25]{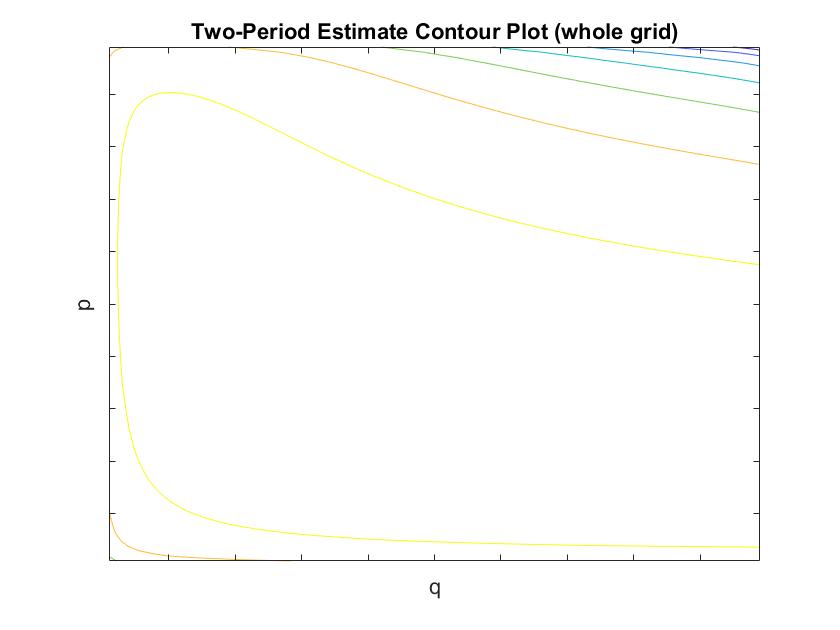}
    \includegraphics[scale=0.25]{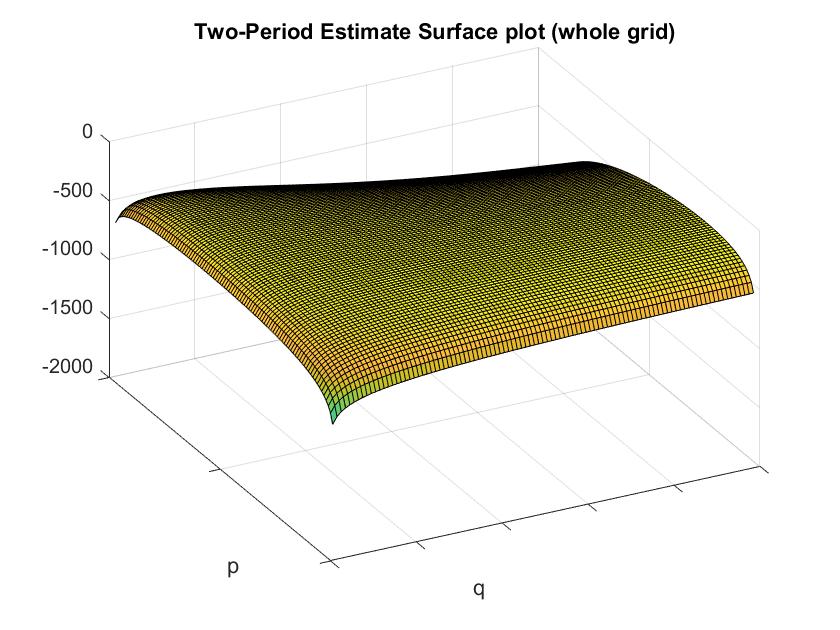}
    \label{fig:my_label}
\end{figure}

\begin{figure}[H]
     \caption{Likelihood Surface and Contour Based around the peak on the Two-Period Estimate from Chapter 2}
    \includegraphics[scale=0.35]{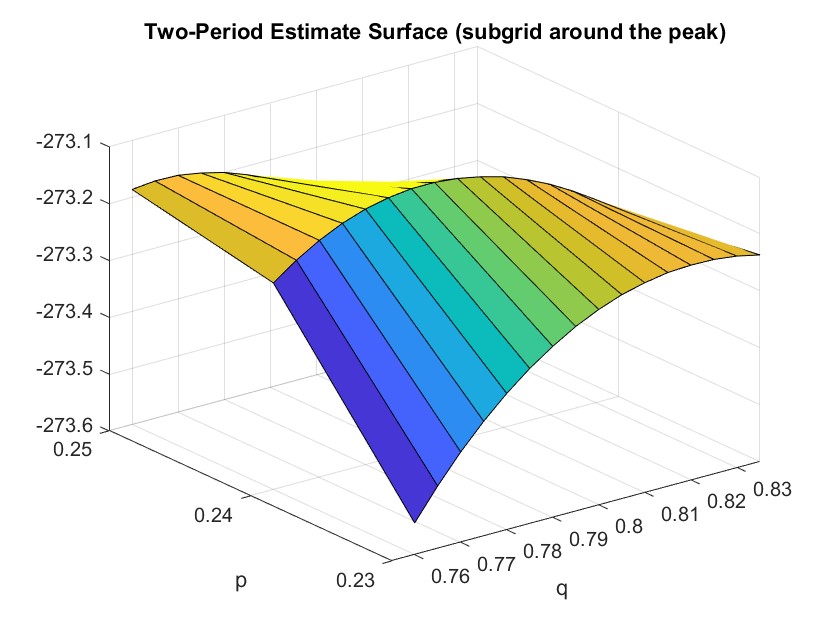}
    \includegraphics[scale=1.25]{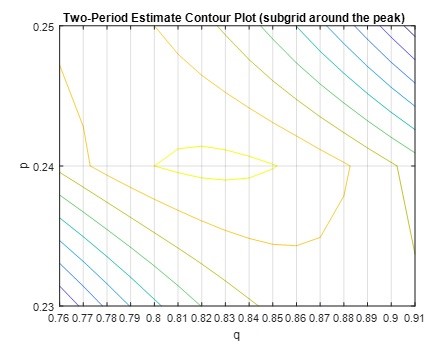}
    \label{fig:my_label}
\end{figure}

\section{Conclusion}
The “Trimming estimator" provides an attractive alternative when exact MLE estimation is infeasible. Substantial increases in computational speed can be achieved whenever the sample comprises a sufficiently large fraction of villages for which no more than one third of the PIIs are trimmed. 
The efficiency gains are substantial, in particular for the diffusion rate estimate. This is in particular the case when the adoption rate is low.  

\section{Appendix}

\subsection{Details on Establishing the (Approximate) Log-likelihood Function}

\begin{itemize}
\item In order to find an expression for the log likelihood function, we will first establish the joint density of $Y$ and $S$ and then integrate out the unobserved information-status vector.
This is an application of the 
{\bf law of total probability}. 
Since $S$ is discrete, we establish the joint probability of the outcome $Y$ and each information scenario that could have generated it and then {\bf sum over information scenarios}. 
\item Conditional on $S_{0:t-1}$, the vector ${\bf Y_t} $ fulfills the {\bf strong Markov property} with respect to its own past values\footnote{That is $Pr(Y_t=y_t|Y_{1:(t-1)}=y_{1:(t-1)},S_{0:t-1}=s_{0:t-1},G,p,q)=
Pr(Y_t=y_t|Y_{(t-1)}=y_{(t-1)},S_{0:(t-1)}=s_{0:t-1},G,p,q)$}
and the {\bf weak Markov property} with respect to the information status vectors
\footnote{That is $Pr(Y_t=y_t|Y_{(t-1)}=y_{(t-1)},S_{0:(t-1)}=s_{0:(t-1)},G,p,q)=
Pr(Y_t=y_t|Y_{(t-1)}=y_{(t-1)},S_{(t-1)}=s_{(t-1)},S_{(t-2)}=s_{(t-2)},G,p,q)$}:  
first, we need to know whether or not the individual already enters the period as a participant (thus we condition on $Y_{t-1}$)
\footnote{If $Y_{i(t-1)}=1$, the individual participates forever, if $Y_{i(t-1)}=0$ and $Y_{it}=1$, she is a new participant. Else, she is a non-participant.}.  
For non-participants, the three possible cases can only be distinguished on the basis of past values of $S$. In particular, when establishing the probability of $Y_t$, we must know  whether or not 
a switch in the information status occurred during the last information exchange (thus we condition on $S_{t-1}$ and $S_{t-2}$).   
\item The individual outcome variables (i.e. the elements of ${\bf Y_t}$) are independent of one another once we condition the outcome of individual $i$ on her own {\bf individual} (scalar) past outcome and her own {\bf individual} (scalar) past two information status variables\footnote{ Consequently, we must condition $Y^i_t$ on $S^i_{t-1}, S^i_{t-2}$ and on $Y^i_{t-1}$}
\item The vector ${\bf S_t} $ fulfills the {\bf strong Markov property} with respect to past values (rows) of both the outcome and the information status.\footnote{$Pr(S_t=s_t|Y_{1:t}=y_{1:t},S_{0:(t-1)}=s_{0:(t-1)},G,p,q)=
Pr(S_t=s_t|Y_{t}=y_t,S_{(t-1)}=s_{(t-1)},G,p,q)
$} This is the case because we need to know whether the outcome signals us that the individual must be informed (since she participates), whether she was previously informed and opted out and what is her probability to be newly informed  
\begin{itemize}
\item It is sufficient to condition on the previous (scalar) {\bf individual outcome } 
\item It is necessary to condition on the entire past {\bf information status vector}: $i's$ probability to be newly informed is a function of her neighbor's information status vectors
\footnote{Consequently, we must condition $S^i_{t}$ on $Y_t^i$ and on $S_{t-1}$}
\end{itemize} 
\end{itemize}
\noindent
These facts can be used to factorize the log likelihood function. \\
\noindent Recall that we denote a possible realization of the random $N \times T-1$ matrix $S$ as an “information scenario" and collect all possible such scenarios in the set $\mathbb{S}$.  
Then $s \in \mathbb{S}$ an information (or, equivalently, a sequence of information status vectors 
$\{ S_1=s_1, S_2=s_2,...,S_{(T-1)}=s_{(T-1)} \} $) that is permissible given the data at hand).
\\ \\
{\bf step 1:} we apply the {\bf law of total probability}
\[ 
Pr \Big( Y_{1:T}=y_{1:T} |s_0, G, p,q \Big)=
Pr \Big( Y_1=y_1, Y_2=y_2,...,Y_T=y_T |S_0=s_0, G, p,q \Big)=\]
\[ =
\displaystyle \sum_{s \in  \mathbb{S}
}
Pr \Big( Y_{1:T}=y_{1:T}, S_{1:(T-1)} \in s| S_0=s_0, G, p,q \Big) =
\]
{\bf step 2:} only the past matters for current outcome variables, only the past and the present matter for current state variables {\bf (assumption 3: timing)} 
 \vspace{.25cm}
 \[
 { \footnotesize
\hspace{-1cm}
 \begin{array}{c}
 Pr\big(Y_1=y_1|G,S_0=s_0,p,q\big) 
\times 
\hspace{3cm} \vspace{.1cm}
\\
\displaystyle \sum_{s \in \mathbb{S} } \Bigg\{
Pr \big(S_{1}=s_{1}\in s| Y_{1}=y_{1},
G,S_0=s_0,p,q\big) 
Pr \big(Y_2=y_2| Y_{1}=y_{1},
 S_{1}=s_1 \in s
,G,S_0=s_0,p,q\big)\times
\\
\displaystyle 
\prod_{t=3}^T 
Pr \big(S_{(t-1)}=s_{(t-1)}\in s| Y_{1:(t-1)}=y_{1:(t-1)},
 S_{1:(t-2)}=s_{1:(t-2)} \in s, 
G,S_0=s_0,p,q\big) \\
Pr \big(Y_t=y_t| Y_{1:(t-1)}=y_{1:(t-1)},
 S_{1:(t-1)}=S_{1:(t-1)} \in s
,G,S_0=s_0,p,q\big) \Bigg\} =
  \end{array}  }
\]
{\bf step 3:} we apply the {\bf Markov properties}
 \vspace{.25cm}
 \[
 {\footnotesize
\hspace{-1cm}
 \begin{array}{c}
 Pr\big(Y_1=y_1|G,S_0=s_0,p,q\big) 
\times 
\hspace{3cm} \vspace{.1cm}
\\
\displaystyle \sum_{s \in \mathbb{S} } \Bigg\{
Pr \big(S_{1}=s_{1}\in s| Y_{1}=y_{1},
G,S_0=s_0,p,q\big) 
Pr \big(Y_2=y_2| Y_{1}=y_{1},
 S_{1}=s_1 \in s
,G,S_0=s_0,p,q\big)\times
\\
\displaystyle 
\prod_{t=3}^T 
Pr \big(S_{(t-1)}=s_{(t-1)}\in s| Y_{(t-1)}=y_{(t-1)},
 S_{(t-2)}=s_{(t-2)} \in s, 
G,S_0=s_0,p,q\big) \\
Pr \big(Y_t=y_t| Y_{(t-1)}=y_{(t-1)},
 S_{(t-1)}=S_{(t-1)} \in s, 
  S_{(t-2)}=S_{(t-2)} \in s
,G,S_0=s_0,p,q\big) \Bigg\} =
  \end{array}  }
\]
{\footnotesize
\begin{align} 
 Pr\big(Y_1=y_1|G,S_0=s_0,p,q\big) 
\times 
\hspace{7cm} 
\notag
\\ 
\displaystyle \sum_{s \in \mathbb{S} } \Bigg\{
\displaystyle 
\prod_{t=2}^T 
Pr \big(S_{(t-1)}=s_{(t-1)}\in s| Y_{(t-1)}=y_{(t-1)},
 S_{(t-2)}=s_{(t-2)} \in s, 
G,S_0=s_0,p,q\big) \notag
\\
Pr \big(Y_t=y_t| Y_{(t-1)}=y_{(t-1)},
 S_{(t-1)}=s_{(t-1)} \in s,
  S_{(t-2)}=s_{(t-2)} \in s
,G,S_0=s_0,p,q\big) \Bigg\} = \notag 
  \end{align}} 

 {\bf step 4:} we apply the {\bf conditional independence across individuals}
\vspace{.15cm}
 \begin{align}\label{finalL}
& \hspace{3cm}  
\displaystyle \prod_{i=1}^N Pr \Big( Y_{i1}=y_{i1} |S_{i0}=s_{i0}, G, p,q \Big)
\times 
\notag
\\ 
& \displaystyle \sum_{s \in \mathbb{S} } \Bigg\{
\displaystyle 
\prod_{t=2}^T 
\Bigg[ \prod_{i=1}^N
 Pr \Big(S_{i(t-1)}=s_{i(t-1)} |Y_{i(t-1)}=y_{i(t-1)}, S_{(t-2)}=s_{(t-2)}\in s,
 S_0=
s_0,G,p,q \Big) \times \notag
\\
&\hspace{2.25cm}
Pr \Big( Y_{it}=y_{it} | Y_{i(t-1)}=y_{i(t-1)},
S_{i(t-2):(t-1)}=s_{i(t-2):(t-1)} \in s,  
S_0=s_0, G, p,q \Big)  \Bigg] \Bigg\} \notag
  \end{align}   

\noindent We have thus established that the likelihood arises as a sum over the product of (scenario specific) probabilities. \\
\noindent
If we number the feasible information scenarios in $\mathbb{S}$ we can rewrite the likelihood function concisely as
\begin{equation} 
= \underbrace{
\underbrace{
\underbrace{
\ell_1 \times 
\displaystyle \sum_{\tau=1}^{\# \mathbb{S} } 
\Delta 
\ell_{2(\tau)} }_{
\substack{\ell_2=
\ell_1 \times \\ Pr(S_1|Y_1,S_0,G,p,q) \\
Pr(Y_2|Y_1,S_1,S_0,G,p,q)}
}
 \Delta
\ell_{3( \tau)}}_{
\substack{
\ell_3=
\ell_2 \times \\
Pr(S_2|Y_2,S_1,G,p,q) \\
Pr(Y_3|Y_2,S_2,S_1,G,p,q)
}
}
 \Delta
\ell_{4( \tau)}}_{
\substack{\ell_4=
\ell_3 \times \\
Pr(S_3|Y_3,S_2,G,p,q) \\
Pr(Y_4|Y_3,S_3,S_2,G,p,q)}
}
\hspace{1cm}
\Delta \ell_{t (\tau)}= \prod_{i=1}^N 
\underbrace{\Delta 
\ell_{it(\tau)}
}_{
\substack{P(Y_{it}=y_{it}|Y_{i(t-1)},S_{i(t-1)}, S_{i(t-2)}) \\
P(S_{i(t-1)}|S_{(t-2)})
}
}
\end{equation} 
Where the latter is the scenario-specific individual contribution to the village log-likelihood function. 
\noindent
From the above, we can see two features that are crucial for the computations: 
\begin{enumerate}
\item The likelihood is established as a sum over scenarios and a product over time periods
and the summands can be computed individually (e.g. by different routines).
The order in which scenarios are evaluated has no importance, which makes it possible for us to “outsource" scenarios, store them 
and process them later. We can have multiple routines (in fact even multiple CPUs) to all add to the same likelihood value for a point on the grid.
\item The Markov property enables us to save intermediate results without bothering about the whole path that lead to them. 
Also, once a sequence has reached the end of the time horizon and added its contribution to the log likelihood function, the respective state vectors are not needed any more and thus memory can be freed to deal with previously saved vectors. 
\end{enumerate} 
As stated above, in any given period, villagers can be categorized into distinct groups: 
\begin{itemize}
\item {\bf former participant: ${\bf Y_{i(t-1)}=Y_{it}=1} $}
\item {\bf new participant: ${\bf Y_{i(t-1)}=0,Y_{it}=1}$ }
\item {\bf non-participant out of reach of the information $ {\bf Y_{it}=0, r_{i(t-1)}=0}  $}
\item {\bf potentially informed individuals (PIIs) $ {\bf Y_{it}=0, r_{i(t-1)}>0}  $}
\end{itemize}
To obtain the individual and scenario specific likelihood contributions, we can simply plug the individual and scenario specific outcome and information status variables into the algorithm we wrote for this purpose. \\ \\ \noindent
{ \bf Density of the outcome variable} \\  \noindent
For the first outcome, the information status is observed.
\[ P(Y_{i1} = y_{i1}|S_{i0}=s_{i0})= s_{i0} \big(y_{i1} p+(1-y_{i1})(1-p) \big)
+(1-s_{i0})(1-y_{i1}) \]
For all subsequent outcomes, the information status is unobserved.
\[ P(Y_{it}=y_{it}| Y_{i(t-1)}=y_{i(t-1)}, S_{i(t-1)}=s_{i(t-1)},
S_{i(t-2)}=s_{i(t-2)},S_0=s_0,G,p,q)= \]

\[ \underbrace{y_{it-1}y_{it} }_{\mbox{ \footnotesize \it 
\shortstack{a previous participant \\ participates for sure}}}   \hspace{0,1cm}+ \hspace{0,1cm}
\underbrace{(1-y_{i(t-1)})y_{it}}_{\mbox{ \footnotesize \it a new participant }} 
\Big[ \underbrace{s_{i(t-1)} (1-s_{i(t-2)})}_{\mbox{\footnotesize \it 
\shortstack{must have been \\newly informed}}} 
\underbrace{p}_{\mbox{\footnotesize \it 
\shortstack{and \\ opted in }
}  }
\Big] +
\]
\[ \underbrace{(1-y_{i(t-1)})(1-y_{it})}_{\mbox{ 
\footnotesize \it a non participant}}
\Bigg[ \underbrace{s_{i(t-1)} (1-s_{i(t-2)})}_{\mbox{  
\footnotesize \it \shortstack{may have been \\newly informed }}}
\underbrace{(1-p)}_{\mbox{ \footnotesize \it 
\shortstack{ and \\ opted out}}}
+ \underbrace{(1-s_{i(t-1)})(1-s_{i(t-2)})}_{
\mbox{ \footnotesize \it may be uninformed }} + \] \[
\underbrace{s^i_{t-2}}_{\mbox{ \footnotesize \it \shortstack{
or may have been informed \\ (and opted out)
in a previous period
}  }}
\Bigg]  \]
\[ 
= y_{i(t-1)}y_{it} +  (1-y_{i(t-1)})y_{it} \Big[s_{i(t-1)} (1-s_{i(t-2)})p\Big] 
+\\
(1-y_{i(t-1)})(1-y_{it}) \Big[1 - s_{i(t-1)} (1-s_{i(t-2)})p \Big] 
\]
Observe that an individual that participated previously will always participate for sure. Further, a switch in the outcome $y_{it}=1 \cap y_{i(t-1)}=0$ indicates that the individual must have been newly informed. Finally, the likelihood to observe no switch is one minus the likelihood to observe a switch. \\
\\
{\bf Density of the information variable} \\  \noindent
\noindent
Note that it follows directly from the timing assumption that the random variable $S_{it}$ depends only on (all individuals') past information status and on the individual's own past participation status.  
\[ P(S_{it}=s_{it}| S_{t-1}=s_{t-1}, Y_{it}=y_{it})=  P(S_{it}=s_{it}| r_{it}, S_{i(t-1)}=s_{i(t-1)},  Y_{it}= y_{it})= \]
\[\underbrace{y_{it} s_{it}}_{ \mbox{ \footnotesize \it 
\shortstack{a participant \\must be\\informed}} }
+ \underbrace{(1-y_{it})}_{ \mbox{ \footnotesize \it 
\shortstack{a non-\\participant}} }
\Bigg[
\underbrace{s_{it}(1-s_{i(t-1)})r_{it}}_{
\mbox{ \footnotesize \it 
\shortstack{may have entered the \\period uninformed\\ 
and newly received\\the information}}}
+ 
\underbrace{(1-s_{it})(1-s_{it-1})(1-r_{it})}_{
\mbox{ \footnotesize \it 
\shortstack{may have entered the \\ period uninformed\\ 
and not received\\the information}}}
+ \] \[
\underbrace{s_{it} s_{i(t-1)}}_{
\mbox{ \footnotesize \it 
\shortstack{may have entered the period informed\\ 
(i.e. previously opted out)}}}
\Big]  \]
\noindent
The probability to observe the data is then the density of the outcome, conditional on the information status, times the density of the information status. Plugging in the concrete values of $y_{it}, y_{i(t-1)}, s_{it}, s_{i(t-1)}$ of the five groups above, straightforwardly leads to the expressions from above.\\ \noindent
Summing up, we can formulate the village log likelihood function in terms of the observed data. Due to the fact that the individual information status is unobserved, we need to make usage of the law of total probability. 
While for the individual outcome, it is sufficient to condition on the past individual participation and information variables, the density of the diffusion variables can only be established by conditioning on the whole system of past state variables for all individuals.
\\
\noindent An information scenario consists of a sequence of information status vectors that, together with the outcome vectors, can be plugged into our algorithm to yield the scenario specific likelihood contribution. 
The difference in establishing the full versus the approximate log likelihood function simply amounts to restricting the number of such scenarios considered.  


\subsection{Performance of the Algorithm}

To ease notation in the following we use $P(y_t)$ to signify $P(Y_t=y_t)$ and $P(s_t)$ to signify $P(S_t=s_t)$. These are the probabilities of a particular outcome pattern or information status, respectively. \\
\noindent
With three time periods and the information initiation being observed, we need to model two information exchanges. 
We dispose over an algorithm that, given one particular outcome (vector $S_1=s_1$) resulting from the first information exchange can compute all possible outcomes of the second information exchange and evaluate the log likelihood function for period $T=3$ accordingly.
If there are $e_1$ Potentially-Informed-Individuals (PIIs) in period one,  the problem arises from the fact that  there are $2^{e_1}$ such realizations of the random vector $S_1$, running the program that many times is prohibitively slow. 
The aim is thus to establish an approximate log likelihood function that considers only a subset $ \mathbb{S}_{1,c} \subset \mathbb{S}_1$ and neglects the remaining scenarios. The number of scenarios considered is $2^d$. The challenge is to pick the set $\mathbb{S}_{1,c}$ such that  the probability mass of all scenarios associated with this period one information scenario is larger than for any scenario considered ($s_{1,c} \in \mathbb{S}_{1,c} $) than for any scenario not considered ($s_{1,nc} \notin \mathbb{S}_{1,c}$). 
Using $\mathbb{S}_2(s_1)\subset \mathbb{S}_2$ to denote the collection of all period two information status vectors that can follow a specific information status vector $s_1$ (i.e. $P(s_1, s_2|y_{1:2},s_0,G,\theta)>0$, we wish all of our selected $s_{1,c}$ to satisfy
\begin{equation}
\label{selection}
   \sum_{
   \substack{s_2 \in \\ \mathbb{S}_2(s_{1,c})}}
P \big(y, s_{1c}, s_2 |s_0,\theta, G) \geq 
\sum_{
\substack{s_2 \in \\ \mathbb{S}_2(s_{1,nc})}
}
P \big(y, s_{1nc}, s_2 |s_0,\theta, G)
\end{equation}
\[ 
\forall s_{1,c} \in \mathbb{S}_{1,c}, s_{1,nc} \notin \mathbb{S}_{1,c} \]
\noindent
 To see why this is important consider 
 \[  \ell =
\sum_{s_{1:2} \in \mathbb{S}}
P \big(y_{1:3}, s_{1:2} |s_0, \theta, G)=
\sum_{s_{1} \in \mathbb{S}_1}
\sum_{
\substack{s_{2} \in \\ \mathbb{S}_2(s_1)}
}
P \big(y_{1:3}, s_{1},s_2 |s_0,
\theta, G)=
 \]
 \begin{equation} \label{selectionn}
 P \big(y_1| s_0, \theta, G)
\sum_{s_{1} \in \mathbb{S}_1}
P \big(s_{1}| y_1, s_0, \theta, G)
P \big(y_2| s_{1},y_1, s_0, \theta, G) \times 
\end{equation}
\[
\sum_{
\substack{s_{2} \in \\ \mathbb{S}_2(s_1)}
}
P \big(s_{2}| y_{1:2},s_1, s_0, \theta, G)
P \big(y_3| y_{1:2},s_2,s_{1},s_0, \theta, G) =
 \]
\[ \sum_{s_{1} \in \mathbb{S}_{1,c}}
\sum_{
\substack{s_{2} \in \\ \mathbb{S}_2(s_{1})
}}
P \big(y_{1:3}, s_{1,}, s_2 |
s_0,\theta, G)+
\sum_{s_{1} \in \mathbb{S}_{1,nc}}
\sum_{
\substack{s_{2} \in \\ \mathbb{S}(s_{1})}
}
P \big(y_{1:3} s_{1}, s_2 |
s_0,\theta, G)=
\widehat{\ell}+ \varepsilon
\]
Let $\underbar{s}_{1,c}$ denote the information status vector in $\mathbb{S}_{1,c}$ for which the LHS of \eqref{selection} is minimal, then, if \eqref{selection} holds, we can quantify the approximation error as 
\[ \varepsilon \leq
 2^{e_1-d}
 \sum_{
\substack{s_{2} \in \\ \mathbb{S}_2(\underbar{s}_{1,c})}
 }
P \big(Y_{1:3}, \underbar{s}_{1,c}
, s_2 |s_0,
\theta, G) \] 
For participants, the information status is unequivocal. The dimensionality problem arises from the potentially informed individuals (PIIs): sufficiently close to have received the information, they do not (by participating) indicate that they have done so. For PIIs, the scenario A: “informed and opted out" and the scenario B: “uninformed" are observationally equivalent. \\
\noindent
As we shall see, the algorithm can lead to choices violating \eqref{selection} due to the fact that it does not anticipate correctly the last part of \eqref{selectionn} and as such, the error bound property fails to hold for some areas of the grid.

\subsection{Erroneous Choices}

\begin{figure}[ht]
\caption{\bf Miniature Village Network}
\label{errch}
\centering \small
\begin{tikzpicture}[scale=.74]
\path 
(0,-0.5) node (ip1) [shape=circle,draw, inner sep=2,outer sep=.5] {ip1 }
(0,-1.5) node (ip2) [shape=circle,draw, inner sep=2,outer sep=0.5] {ip2 }
(0,-3) node (ip3) [shape=circle,draw, inner sep=2,outer sep=0.5] {ip3 }
(0,-4) node (ip4) [shape=circle,draw, inner sep=2,outer sep=0.5] {ip4}
(0,-5) node (ip5) [shape=circle,draw, inner sep=2,outer sep=0.5] {ip5 }
(2,-1) node (6) [shape=circle,draw, inner sep=2,outer sep=0.5] {6 }
(2,-3) node (7) [shape=circle,draw, inner sep=2,outer sep=0.5] {7 }
(2,-4) node (8) [shape=circle,draw, inner sep=2,outer sep=0.5] {8 }
(2,-5) node (9) [shape=circle,draw, inner sep=2,outer sep=0.5] {9 }
(4,1) node (10) [shape=circle,draw, inner sep=2,outer sep=.5] {10}
(4,0) node (11) [shape=circle,draw, inner sep=2,outer sep=.5] {11 }
(4,-1) node (12) [shape=circle,draw, inner sep=2,outer sep=0.5] {12 }
(4,-2) node (13) [shape=circle,draw, inner sep=2,outer sep=0.5] {13 }
(4,-4) node (14)
[shape=circle,draw, inner sep=2,outer sep=0.5] {14 }
;
\draw [dashed] (ip1)--(6);
\draw [dashed] (ip2)--(6);
\draw [dashed] (ip3)--(7);
\draw [dashed] (ip4)--(8);
\draw [dashed] (ip5)--(9);
\draw [dashed] (10)--(6);
\draw [dashed] (11)--(6);
\draw [dashed] (12)--(6);
\draw [dashed] (13)--(6);
\draw [dashed] (14)--(8);
\draw [dashed] (14)--(9);
\draw [dashed] (14)--(7);
\end{tikzpicture}
\end{figure}
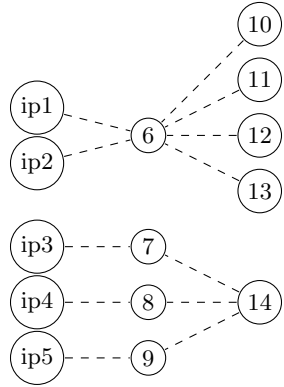

\noindent Consider the following example.
\\
\noindent
Define a sub-graph as consisting of a final agent $k \in e_2$ and the collection of intermediate agents ($j=1,..,J \in e_1$) that are situated on the indirect paths from $k$ to one or several IPs. Figure \ref{errch} depicts two independent sub-graphs.

\noindent
Assume that individuals 6, 7, 8, 9 are subject to trimming. If $A_6>B_6$ we will trim 6 to A. However, it can be that  
\begin{equation}
\label{ex1}
    A_6 
    \prod_{i=10}^{13} (A_i+B_i)< 
     (B_6) (A_6+B_6)
\end{equation}
(this is for instance the case for $p=0.5=q$).  
In this case, setting 6 to scenario A turns out to be a sub-optimal choice. \\
\noindent
By the same argument if $A_7<B_7$, it may be that 
\begin{equation}
\label{ex2}
    (\prod_{i=7}^9 A_i) (A_{14}+B_{14})>
    \prod_{i=7}^9 B_i(A_i+B_i)
\end{equation}
\noindent
(This is for instance the case for $p=0.4, q=0.6$). \\ \noindent
The simple algorithm thus results in \eqref{ex1} and \eqref{ex2} violating \eqref{selection}.
\noindent
To evaluate the performance of the algorithm, it is informative to understand when applying it entails mistakes. We define a mistake as a choice that violates such that there exists a different choice in the current period that guarantees that a larger fraction of probability mass is covered in the final period.  \\
Note that $A_i+B_i=1-pr_i$ is decreasing in $r_i$. Consequently, if we choose scenarios that imply that intermediate PIIs are informed, then these should c.p.\ be the ones with the largest number of IP friends: they have a high $A_i$ and a low $B_i(A_i+B_i)$. 
Furthermore when an intermediate agent is informed in the first exchange, she can inform her neighbours in the second exchange. Say individual $i$ has $\omega_i$ neighbours, then having $i$ informed in period one leads to 
$A_i \prod_{j=1}^{\omega_i} (A_j+B_j)$ as this decreases in $\omega_i$, thus if we choose scenarios that imply that intermediate agents are informed, these should c.p.\ be the ones with the smallest out-degree. However, final agents are regularly linked to not one, but several intermediate PIIs. To account for the the total impact a PII has on the final agents, we need to consider her “ip-betweenness". We define the latter as the sum of the fractions of links between an ip and a final agent that pass by the intermediate agent. A high  ip-betweenness signifies that the intermediate agent is linked to many final agents or that these final agents she is linked with are linked to few other intermediate agents. An ip-betweenness smaller than one indicates that the final agents she is linked to are also a neighbours of many other intermediate agents. \\
\noindent 
There are two situation that may result in erroneous choices. \\
\\
\noindent
{\bf Mistake 1: betweenness exceeds the in degree and $\bf A_i>B_i$} \\
For facilitate the exposition, we first assume that all final agents are homogeneous and only linked to one intermediate agent and thereafter give a formula that is valid in general. This case is the first example from the above.
Then we have 
\begin{align*}
A_i \left(\prod_j 
(A_j+B_j)\right) &= 
B_i (A_i+B_i)\\
\frac{A_i}{B_i} &= 
\frac{(A_i+B_i)}{
(1-pq)^b
} \\ \vspace{.5cm}
b \rightarrow \infty 
\hspace{.75cm}
\frac{A_i}{B_i} &< 
\underbrace{\frac{(A_i+B_i)}{
(1-pq)^b}}_{\rightarrow \infty}
\end{align*}
If we let $b$ go to infinity while holding the in degree constant, the denominator of the second term goes to zero, implying that the second term becomes larger than the first. This means that “$i$ is uninformed in the first exchange" ultimately results in a larger probability mass then “$i$ is informed in the first exchange" despite the fact that $A_i>B_i$ and as such, the choice was incorrect. \\
\\ \noindent
{\bf Mistake 2: the in degree exceeds the betweenness and $\bf A_i<B_i$} \\
\noindent
For clarity, assume that intermediate agents are homogeneous and linked to one final agent, i.e.\ they each have a share of $b<1$ of the paths from IPs to the final agent that pass through them and as a consequence, there are $1/b$ intermediate agents. This is the situation in the second example above. 
Since $B_i>A_i$, the algorithm trims the intermediate agents to B. This is the optimal choice if 
\begin{align*}
 \prod_i A_i  
(A_j+B_j) &<
 \prod_i B_i (A_i+B_i)
 \\
 \Big(
\frac{
A_i}{B_i}\Big)^{1/b} &<
\frac{\Big(A_i+B_i\Big)^{1/b}
}{
(A_j+B_j)
}
\\ \vspace{.5cm}
\frac{A_i}{B_i} &<
\frac{(A_i+B_i)}{ 
(1-p(1-(1-q)^{1/b})^b
}\end{align*}
However, if the number of intermediate agents is increased, $b$ becomes very small and as a consequence, the denominator of the RHS approaches one, such that the RHS expression converges to $A_i+B_i<1$. If we are at a point on the grid where the difference between $A_i$ and $B_i$ is small, the LHS is close to one and thus exceeds the RHS, such that we conclude that the choice has been erroneous, i.e.\
\[ b \rightarrow 0 
\hspace{.75cm}
\underbrace{\frac{A_i}{B_i}}_{\approx 1} > 
\underbrace{\frac{(A_i+B_i)}{
(1-p(1-(1-q)^{1/b})^b}}_{\rightarrow A_i+B_i<1 } \]
When $b$ is small, setting all intermediate agents to “informed" in the first exchange implies a higher reception rate for the final agent, thus decreasing $A_j+B_j$ and making this choice less attractive. However, $A_j+B_j$ decreases less than proportionally and as the number of intermediate agents grows large, the further  decrease becomes negligible. Setting all intermediate agents to “uninformed" features a zero reception rate for the final agent but, on the other hand, implies that we need to consider two scenarios in the next period for each intermediate agent, which decreases the probability mass. When $B_i>A_i$ induces us to set intermediate agents to “uninformed", we neglect this future impact, which may result in a sub-optimal choice.

\noindent
The formulas above use the easier cases in which either final or  intermediate agents are homogeneous. More generally, however, intermediate agents can have different link portfolios and sub-graphs can overlap. Now, there is no straightforward relationship between $b$ and the number of agents in the sub-graph, however, the two are still proportional. A lower b value implies that a agent is liked to few, or densely connected final agents. A higher b value on the other hand signifies that she is one of the few neighbours of a final agent, or linked to a large number of them. 
Assume that there are $J+1$ intermediate agents who are linked to $K$ final agents. To focus on the last intermediate agent, we assume that the equation holds strictly for the $J$ remaining ones, implying 
\begin{align*}
    \label{ESA2}
    (\prod_{i=1}^J A_i) 
    \prod_{k=1}^K
    \big[A_{k(J)}+B_{k(J)} \big] &=
    \prod_{i=1}^{J} B_i (A_i+B_i)
\\
    \prod_{i=1}^J 
    \frac{A^i}{B^i} &=
    \frac{\prod_{i=1}^{J}  (A_i+B_i)}{\prod_{k=1}^K
    \big[A_{k(J)}+B_{k(J)}\big]}
\end{align*}
If we add the last agent, this becomes 
\begin{equation*}
    \prod_{i=1}^J 
    \frac{A_i}{B_i}
    \frac{A_l}{B_l}
    =
    \frac{\prod_{i=1}^{J}  (A_i+B_i)}{\prod_{k=1}^K
    \big[A_{k(J)}+B_{k(J)}\big]}
     \frac{\prod_{k=1}^K \big[A_{k(J)}+B_{k(J)}\big]}{\prod_{k=1}^K
    \big[A_{k(J+1)}+B_{k(J+1)}\big]} (A_l+B_l)
\end{equation*}
The second term reflects the fact that now all final agents have one more potential linking partners. We cancel out the equality to obtain 
\begin{equation*}
    \frac{A_l}{B_l}
    =
     \frac{\prod_{k=1}^K \big[ A_{k(J)}+B_{k(J)}\big]}{\prod_{k=1}^K
    \big[A_{k(J+1)}+B_{k(J+1)}\big]} (A_l+B_l)
\end{equation*}
Further, in the first term on the RHS, we can cancel out all agents that are not linked to agent $l$ and note that the remaining $\tilde{K}$ ones gain one additional link when $l$ is added
\begin{equation*}
    \frac{A_l}{B_l}
    =
     \frac{\prod_{k=1}^{\tilde{K}} \big[ 
     1-p(1-(1-q)^{x^k}
     \big]}{\prod_{k=1}^{\tilde{K}}
    \big[
    1-p(1-(1-q)^{x^k+1}
    \big]} (A_l+B_l)
\end{equation*}
The equality may fail to hold. If $x^k$ is small, then $l$ is the only or one of the few neighbours for many final agents and the first term on the RHS is larger than one. If there are many such final agents $\tilde{K}$ is large thus counteracting the multiplication by $A_l+B_l$ ad the entire RHS exceeds one. Then $l$ has a substantial impact on many final agents ad despite the fact that $A_l>B_l$, the optimal choice is to trim her to $B_l$.
\begin{equation*}
\tilde{K} \rightarrow K,
\hspace{.5cm}
x^k \rightarrow 0, 
\hspace{.5cm}
    \frac{A_l}{B_l}
    <
     \frac{(A_l+B_l)}{(1-pq)^{\tilde{K}}}
\end{equation*}
which implies committing error one. 
\\ \noindent
If on the other hand $x_k$ is large and/or $\tilde{K}$ is small, then the first term on the RHS is close to one. Agent $l$ has a negligible impact on the others and the condition becomes
\begin{equation*}
\tilde{K} \rightarrow 0,
\hspace{.5cm}
x^k \rightarrow \infty,  
\hspace{.5cm}
    \frac{A_l}{B_l}
    > A_l+B_l
\end{equation*}
If $B_l>A_l$ and the difference is sufficiently small, we end up committing error two.

\subsection{The Error Function}

We have seen above that, by the law of total probability, the likelihood function arises as a sum, the summands being the (joint) probability of the data and each (unobserved) information status matrix that could have generated it. We have also seen that in dense Networks with low participation rates, the total number of possible information scenarios (and thus the number of summands of the log likelihood function) easily explodes. We have proposed to tackle this problem by neglecting highly unlikely scenarios. Consequently, our method leads to an approximation error. We now quantify the approximation error, analyze the systematic pattern it exhibits and outlay how this pattern can be used for identifying the maximum of the village Log-likelihood function.  
\\ \noindent 
As before, we  let $\mathbb{S}$ denote the set of all possible sequences of information status vectors $S_1,...,S_{T-1}$ that are conform with the observed outcome matrix $Y_{(1:T)}$. If $s$ stands for a particular sequence in this set then the log likelihood function of village $v$ is given by 
\begin{align}
\mathcal{L}_v =
\log \Bigg(
\displaystyle \sum_{s \in \mathbb{S}_v } 
Pr (Y_{(1:T),v}=y_{(1:T),v},s|s_{0,v},G_v,p,q ) \Bigg) \notag
\end{align}

\noindent We have also noted that the curse of dimensionality arises from the typically large size of “potentially informed individuals" (PIIs): with two possible information statuses, 
each additional non-participant 
that comes in reach of the information in period 1 (respectively, 2, 3) can potentially quadruple (respectively, triple, double) the possible realizations of the random matrix $S$.
We need to reduce the number of sumamnds of the log likelihood function and we achieve this reduction by 
limiting the set of PIIs for which both scenarios (“informed and opted out" and “uninformed") are considered. 
For each point on the grid, our algorithm can foresee the ordering of information scenarios according to their probabilities to occur even without having to actually compute (and store) these probabilities. Having established the ordering of scenarios in $\mathbb{S}$, we can select as many scenarios as our storage and computing capacities allow, knowing we selected the most likely ones.  
As a consequence,  
whenever a specific information scenario is highly unlikely, it will be neglected and the PIIs involved in this unlikely scenario will be pushed to their more likely default value.   \\
\noindent 
Since out of the set $\mathbb{S}_v$ we are only considering a limited number of elements. Partitioning $\mathbb{S}=\mathbb{S}_{c,v}  \cup \mathbb{S}_{nc,v}  $ (where the subscript c (nc) stands for the considered (not considered) scenarios such that $\mathbb{S}_{c,v}  \cap \mathbb{S}_{nc,v}  = \emptyset $). We can write 
\begin{align} \mathcal{L}_v = 
\log
\Bigg(
\displaystyle \sum_{s \in \mathbb{S}_{c,v} } 
&
Pr (Y_{(1:T),v}=y_{(1:T),v},s|s_{0,v},G_v,p,q)  +  \notag \\
& \hspace{-1cm}
\displaystyle \sum_{s \in 
\mathbb{S}_{nc,v} } 
Pr (Y_{(1:T),v}=y_{(1:T),v},s|s_{0,v},G_v,p,q) \Bigg) \notag
\end{align}
\noindent In the following, $\hat{\ell}$ ($\hat{\mathcal{L} }$) denotes the approximate (log) likelihood function. 
\[ \mathcal{L}_v = 
 \log \bigg(
\widehat{ \ell_v}+ 
\ell_v - \widehat{ \ell_v}
\bigg)
=
\log \Bigg(
 \widehat{\ell_v}
 \Bigg[
1+ \frac{
\ell_v - \widehat{ \ell_v}
}{\widehat{\ell_v}} 
\Bigg] \Bigg)
=
 \log \big(
 \widehat{\ell_v} \big) +
 \log \Bigg(
1+ \frac{
\ell_v - \widehat{ \ell_v}
}{\widehat{\ell_v}} 
\Bigg)=
\]
\begin{equation}\label{Errordef}
    \mathcal{L}_v=
\widehat{
\mathcal{L}_v}+
\varepsilon_v
\end{equation}
Therefore, we have the sample Log-likelihood function 
\[ \mathcal{L} = 
\log \Big( \prod_v \ell_v
\Big) = 
 \sum_v \log (\ell_v)=
 \sum_v \mathcal{L}_v=
 \sum_v \widehat{\mathcal{L}}_v+ 
 \sum_v  \varepsilon_v=
 \widehat{\mathcal{L}}+\varepsilon
\]

\noindent The error function is always piece-wise linear in the trimming value. Furthermore, it is close to being convex if the algorithm does not commit too many errors on average. 
Convexity of the error function is useful because it enables us to get an estimate of the maximal error by interpolation. 
Knowing that at $e^1=d$ the error is zero, we can use the slope $\Delta \varepsilon_1$ to trace back the maximal error level: as the error function is concave, the actual error curve will thus under our interpolated one and hence we obtain a conservative estimate.

\noindent
The error function is piecewise linear in $d$. It can be convex, concave or change curvature over the range of $d=1,...,e^1$. However, as long as the algorithm makes no large mistakes on average, it is close to being convex.

\begin{proof}
Define $\mathbb{S}_{c,d}$ as the set of scenarios considered if $d$ agents are not trimmed. Then define 
$P(s_d)= \sum_{s \in \mathbb{S}_{c,d}}
-\sum_{s \in \mathbb{S}_{c,d-1}}
$ as the additional scenarios considered when the $d^{th}$ agent is not trimmed. 
Since $\mathcal{L}=
\varepsilon_d=
\widehat{\mathcal{L}}_{d}$
thus the change in the error is 
\[  \Delta \varepsilon_d=
\widehat{\mathcal{L}}_{d-1}-
\widehat{\mathcal{L}}_{d1}=
\log \bigg( \sum_{x=1}^{d-1}P(s_x)\bigg)-
\log \bigg( \sum_{x=1}^{d}P(s_x)\bigg)=\]
\[\log \bigg( \sum_{x=1}^{d-1}P(s_x)\bigg)-
\log \bigg( \sum_{x=1}^{d-1}P(s_x)
+P(s_d)\bigg)=
\]
\[\log \bigg( \sum_{x=1}^{d-1}P(s_x)\bigg)-
\log \bigg( \sum_{x=1}^{d-1}P(s_x)
\bigg(1
+\frac{P(s_d)}{\sum_{x=1}^{d-1}P(s_x)}\bigg)
=  \]
\[ \log \bigg( \sum_{x=1}^{d-1}P(s_x)\bigg)-
\log \bigg( \sum_{x=1}^{d-1}P(s_x)\bigg)-
\log \bigg(1
+\frac{P(s_d)}{\sum_{x=1}^{d-1}P(s_x)}
\bigg) = \]
\[ - \log \bigg(
1
+\frac{P(s_d)}{\sum_{x=1}^{d-1}P(s_x)}\bigg)
\]
The error function always slopes downwards. The slope exceeds one if 
$\frac{P(s_d)}{\sum_{x=1}^{d-1}P(s_x)}>1$ that is the sum of all new scenarios is more likely than the sum of all scenarios considered so far, in which case the error curve exhibits a kink. A necessary and sufficient condition for convexity is \[ 
- \log \bigg(
\frac{P(s_{d-1})}{\sum_{x=1}^{d-2}P(s_x)} \bigg) <
-
\log \bigg(
\frac{P(s_d)}{\sum_{x=1}^{d-1}P(s_x)}
\bigg)
\] 
\[ 
\frac{P(s_{d-1})}{\sum_{x=1}^{d-2}P(s_x)}  >
\frac{P(s_d)}{\sum_{x=1}^{d-1}P(s_x)}
\hspace{.5cm}
\forall d
\] 
\end{proof}
\noindent
In the numerator are the new scenarios that arise when $d$ is not trimmed. These are thus the scenarios in which she is set to the smaller of the two choices $A$ or $B$, whereas in the denominator are the same scenarios with PII $d$ being trimmed to her threshold. As such, if the choice is correct, then the slope of the error function is indeed decreases in absolute value and as such, convexity holds. 
Importantly, it is not necessary for convexity that the algorithm commits no error. Any choice to trim to a threshold can be evaluate conditioning on the scenarios all other agents and as such, a choice is rarely erroneous in all of these cases. Convexity holds as long as the fraction of the total probability mass of the non-treshhold scenarios over the total probabilities of the treshold scenarios is smaller than one and keeps decreasing. The error function is close to being convex as long as the algorithm does not commit too many errors on average.

\section{Bibliography}
\nocite{*}
\bibliographystyle{chicago}
\bibliography{my}

\end{document}